\documentclass[12pt,preprint]{aastex}
\usepackage{amsmath,amssymb}
\usepackage{graphicx}
\usepackage{float}
\usepackage{multirow}

\slugcomment{KSUPT-13/2 \hspace{0.5truecm} May 2013}

\begin{document}

\title{Binned Hubble parameter measurements and the cosmological 
deceleration-acceleration transition}

\author{Omer Farooq, Sara Crandall, and Bharat Ratra}

\affil{Department of Physics, Kansas State University, 
                 116 Cardwell Hall, Manhattan, KS 66506, USA} 
\email{omer@phys.ksu.edu, sara1990@k-state.edu, ratra@phys.ksu.edu}

\begin{abstract}

Weighted mean and median statistics techniques are used to combine 23 independent lower redshift,
$z<1.04$, Hubble parameter, $H(z)$, measurements and determine binned forms of $H(z)$. When these are combined
with 5 higher redshift, $1.3\leqslant  z \leqslant 2.3$, $H(z)$ measurements the resulting 
constraints on cosmological parameters, of three cosmological models, that follow from the weighted-mean
binned data are almost identical to those derived from analyses using the 28 independent
$H(z)$ measurements. This is consistent with what is expected if the lower redshift measurements errors 
are Gaussian. Plots of the binned weighted-mean $H(z)/(1+z)$ versus $z$ data are consistent with the
presence of a cosmological deceleration-acceleration transition at redshift $z_{\rm da}=0.74 \pm 0.05$ 
\citep{farooq3}, which is expected in cosmological models with present-epoch energy budget dominated by
dark energy as in the standard spatially-flat $\Lambda$CDM cosmological model.

\end{abstract}

\maketitle
\section{Introduction}
\label{intro}

In the standard cosmological model\footnote{ For recent reviews
see, e.g., \cite{wang2011ab},
\cite{Li2012b}, and \cite{tsujikawa13}. In this paper we assume that general relativity 
provides an adequate description of gravitation on cosmological scales of interest;
for discussions of modified gravity see \cite{Capozziello2011}, 
\cite{trodden12} and references therein.} dark energy dominates the current epoch 
energy budget, but was less important in the past when non-relativistic (cold dark 
and baryonic) matter dominated. The transition from non-relativistic matter 
dominance to dark energy dominance results in a transition from decelerated 
to accelerated cosmological expansion. The existence of this transition is a strong 
prediction of the standard cosmological model and attempts have been made to
measure the transition redshift.\footnote{
See \cite{Lu2011a}, \cite{Giostri2012}, \cite{Lima2012}, and 
references therein.}
 However, only very recently has this become possible, due to high redshift 
(i.e., $z$ above the deceleration-acceleration transition) data that recently became
available, with the most striking being the \cite{busca12} measurement of the 
Hubble parameter $H(z=2.3)=224 \pm 8$ km s$^{-1}$ Mpc$^{-1}$, well in the matter 
dominated epoch of the standard $\Lambda$CDM model. 

From a compilation of 28 independent $H(z)$
measurements over $0.07\leqslant  z \leqslant 2.3$ \citep[][hereafter FR, Table 1]{farooq3}, 
the transition redshift was found to be $z_{\rm da} = 0.74 \pm 0.05$. 
This was determined from the 6 best-fit transition redshifts measured in three 
different cosmological models, $\Lambda$CDM, XCDM, and $\phi$CDM, for two different 
Hubble constant priors.

The spatially-flat $\Lambda$CDM model \citep{peebles84} is the reigning standard 
cosmological model. In this paper we consider the more general $\Lambda$CDM model that 
allows for non-zero space curvature. In the standard model 
the cosmological constant, $\Lambda$, contributes around 70$\%$ of the 
present cosmological energy budget, non-relativistic, pressure-less, cold
dark matter (CDM) contributes a little more than 20$\%$, and non-relativistic baryonic matter 
makes up the remaining 5$\%$ or so. In the $\Lambda$CDM model time-independent dark energy, 
$\Lambda$, is modeled as a spatially homogeneous 
fluid with equation of state $p_{\rm \Lambda} = -\rho_{\rm \Lambda}$ where 
$p_{\rm \Lambda}$ and $\rho_{\rm \Lambda}$ are the fluid pressure and energy 
density respectively. It has been known for a while now that the spatially-flat 
$\Lambda$CDM model is consistent with most observational data.\footnote{For 
early indications see, e.g., \cite{jassal10}, \cite{wilson06}, \cite{Davis2007}, and \cite{allen08}. 
Note, however, there are some preliminary observational hints that the standard CDM 
structure formation model, assumed in the flat $\Lambda$CDM cosmological model, 
might need to be improved upon 
\citep[][and references therein]{Peebles&Ratra2003, Perivolaropoulos2010}.} 
It is also well known that if, instead of staying constant
like $\Lambda$, the dark energy density gradually decreased 
in time (and correspondingly slowly varied in
space), it would alleviate a conceptual coincidence 
problem associated with the $\Lambda$CDM model.\footnote{For recent discussions 
of time-varying dark energy models see 
\cite{Guendelman2012}, \cite{Wang12}, \cite{De-Santiago12}, 
\cite{lima13}, \cite{Capozziello13}, \cite{Adak2012},  
and references therein}

A widely-used parameterization of time-evolving dark energy, XCDM, parameterizes dark energy as 
a spatially-homogeneous time-varying $X$-fluid with equation of state 
$p_X=\omega_X\rho_X$. Here, the equation of state parameter $\omega_X$ can 
take any time-independent value less than $-1/3$. For computational simplicity we consider only
spatially-flat XCDM models. The XCDM parametrization reduces to the flat $\Lambda$CDM 
model for $\omega_X=-1$. For all other values of $\omega_X<-1/3$, the XCDM parametrization 
is incomplete since it does not describe spatial inhomogeneities 
\citep[see, e.g.,][]{ratra91, podariu2000}.

A simple, consistent, and complete model of slowly-varying dark 
energy density is the $\phi$CDM model \citep{Peebles&Ratra1988, 
Ratra&Peebles1988}. Here dark energy is modeled as a scalar 
field, $\phi$, with a gradually decreasing (in $\phi$) potential energy 
density $V(\phi)$. In this paper we assume an inverse-power-law potential energy 
density $V(\phi) \propto \phi^{-\alpha}$, where $\alpha$ is a nonnegative 
constant \citep{Peebles&Ratra1988}. When $\alpha = 0$ the
$\phi$CDM model reduces to the corresponding $\Lambda$CDM case. 
For computational simplicity we again only consider the spatially-flat
cosmological case for $\phi$CDM.

In addition to being affected by the cosmological model used in the analysis, the 
measured deceleration-acceleration transition redshift $z_{\rm da}$ depends on the assumed 
value of the Hubble constant. Consequently, to quantify the effect, we use
two Gaussian $H_0$ priors in the analyses. The first prior is 
$\overline{H}_{0}$ $\pm$ $\sigma_{H_{0}}$ = 68 $\pm$ 2.8 km s$^{-1}$ Mpc$^{-1}$. 
This comes from a median statistics analysis of 553 $H_0$ measurements \citep{Chen2011a} 
and is consistent with the earlier estimates of \cite{Gott2001} and \cite{Chen2003}. The
second prior of  
$\overline{H}_{0}$ $\pm$ $\sigma_{H_{0}}$ = 73.8 $\pm$ 2.4 km s$^{-1}$ Mpc$^{-1}$ comes 
from recent Hubble Space Telescope measurements \citep{Riess2011}.\footnote{Other
recent measurements are consistent with either the smaller or larger $H_0$ value
we consider, see, e.g., \cite{Freedman12}, \cite{Sorce12},
and \cite{Tammann12}, although it might now be significant that both BAO 
\citep[see, e.g.,][]{Colless12} and \textit{Planck} CMB anisotropy \citep{Ade13} 
measurements favor the lower $H_0$ value we use. It might also be significant that the
lower value of $H_0$ does not require the presence of dark radiation
\citep[][and references there in]{calabrese12}.}

In FR we determined the redshift of the deceleration-acceleration
transition by finding the mean and standard deviation of the six best-fit $z_{\rm da}$ values
in the 3 models (with 2 different $H_0$ priors). Here we use a different technique to measure
 $z_{\rm da}$ and the related uncertainty in each of these six cases. We then determine 
 summary estimates of $z_{\rm da}$ by considering various weighted mean combinations of 
 these six estimates. The transition redshifts take the forms
\begin{align}
&z_{\rm da}=\left({\frac{2 \Omega_{\Lambda}}{\Omega_{m0}}}\right)^{1/3}-1,\\
&z_{\rm da}=\left({\frac{\Omega_{m0}}{(\Omega_{m0}-1)(1+3 \omega_X)}}\right)^{1/{3\omega_{X}}}-1,
\end{align}
for the $\Lambda$CDM and XCDM cases where $\Omega_{\Lambda}$ and $\Omega_{m0}$ 
are the cosmological constant and non-relativistic matter density parameters. 
As for $\phi$CDM, from Eqs.\ (3) of \cite{Peebles&Ratra1988} we first derive
\begin{eqnarray}
\frac{\ddot{a}}{a}&=&-\frac{4 \pi G}{3} \left[\rho_{m}+\rho_{\phi}(1+3\omega_{\phi})\right] \nonumber \\
				&=& -\frac{1}{2} H_0^2 \left[\Omega_{m0} (1+z)^3+\Omega_{\phi}(z,\alpha)(1+3\omega_{\phi}(z))\right],
\label{eq:friedman}
\end{eqnarray}
where $\Omega_{\phi}(z)$ is the scalar field energy density parameter and
\begin{align}
\omega_{\phi}(z)=\frac{\frac{1}{2}\dot{\phi^2}-V(\phi)}{\frac{1}{2}\dot{\phi^2}+V(\phi)}.
\end{align}
The redshift $z_{\rm da}$ is determined by requiring that the right hand side 
of Eq.\ (\ref{eq:friedman}) vanish,
\begin{align}
&\Omega_{m0} (1+z_{\rm da})^3+\Omega_{\phi}(z_{\rm da},\alpha)\left[1+3~\omega_{\phi}(z_{\rm da})\right]=0.
\label{eq:zdaphi}
\end{align}
To determine $z_{\rm da}$ we numerically integrate the $\phi$CDM 
model equations of motion, Eqs. (3) of \cite{Peebles&Ratra1988},
using the initial conditions described there. These solutions 
determine the needed functions in Eq. (\ref{eq:zdaphi}), which we then numerically
solve for $z_{\rm da}(\Omega_{m0},\alpha)$. 

To find the expected values
$\langle z_{\rm da} \rangle$ and $\langle z_{\rm da}^2 \rangle$ 
we use
\begin{eqnarray}
\label{expected zda}
\langle z_{\rm da} \rangle=\frac{\iint z_{\rm da}(\textbf{p}) \mathcal{L}(\textbf{p}) d\textbf{p}}{\iint \mathcal{L}(\textbf{p}) d\textbf{p}},~~~~~~~~~~~~
\langle z_{\rm da}^2 \rangle=\frac{\iint z_{\rm da}^2(\textbf{p}) \mathcal{L}(\textbf{p}) d\textbf{p}}{\iint \mathcal{L}(\textbf{p}) d\textbf{p}}.
\end{eqnarray} 
Here $\mathcal{L}(\textbf{p})$ is the $H(z)$ data likelihood function after marginalization
over the $H_0$ prior in the model under consideration. It depends only on the model 
parameters $\textbf{p}=(\Omega_{m0},\Omega_{\Lambda})$ for $\Lambda$CDM,
$=(\Omega_{m0},\omega_{X})$ for XCDM, and 
$=(\Omega_{m0},\alpha)$ for $\phi$CDM. 
The standard deviation in $z_{\rm da}$ is calculated from the standard formula
$\sigma_{z_{\rm da}}=\sqrt{\langle z_{\rm da}^2 \rangle-\langle z_{\rm da} \rangle^2}$.
The results of this computation are summarized in Table 1.

It is reassuring that the results of the penultimate and the last columns of Table 1 are 
very consistent. FR determined a summary estimate of $z_{\rm da}=0.74 \pm 0.05$ by 
computing the mean and standard deviation of the six values in the last column of Table 1. It is 
of interest to estimate similar summary values for each of the two $H_{0}$ priors. We find that 
$z_{\rm da}=0.70 \pm 0.05$ ($z_{\rm da}= 0.77 \pm 0.04$) for 
$H_{0}\pm\sigma_{H_0}$ = 68 $\pm$ 2.8 (73.8 $\pm$ 2.4) km s$^{-1}$ Mpc$^{-1}$. 
Perhaps  more realistic summary estimates are determined by the weighted means
of the two sets of 3 values in the penultimate column of Table 1: 
$z_{\rm da}=0.69 \pm 0.06$ ($z_{\rm da}=0.76 \pm 0.05$) for 
$H_{0}\pm\sigma_{H_0}$ = 68 $\pm$ 2.8 (73.8 $\pm$ 2.4) km s$^{-1}$ Mpc$^{-1}$, 
and $z_{\rm da}=0.74 \pm 0.04$ is the result if all six values are used.

More conventionally, cosmological data are used to constrain model parameters 
values such as $\Omega_{m0}$ and $\Omega_{\Lambda}$ for the $\Lambda$CDM model. 
A number of different data sets have been used for this purpose. 
These include Type Ia supernova (SNIa) apparent magnitude verses redshift data \citep[e.g.,][]{Ruiz2012,
chiba2013, cardenas2011, Liao2013, Farooq2012, campbell13}, cosmic microwave 
background (CMB) anisotropy measurements \citep[][and references therein]{Ade13}, 
baryonic acoustic oscillation (BAO) peak length scale data \citep[][and references therein]{Mehta2012,
Anderson12,  Li2012a, Scovacricchi2012, Farooq2013a}, galaxy cluster gas mass fraction
as a function of redshift \citep[e.g.,][]{allen08, Samushia&Ratra2008, 
tongnoh11, Lu2011b, solano12, landry12}, and, of special interest here, measurement of the Hubble
parameter as a function of redshift \citep[][and references therein]{Jimenezetal2003, 
Samushia&Ratra2006, samushia07, Sen&Scherrer2008, Chen2011b, Aviles2012, 
wangjcap12, campos12, Chimento13}. These data,
separately and in combination, provide strong evidence for accelerated cosmological
expansion at the current epoch.\footnote{Other data, with larger error
bars, support these results. See, e.g., \cite{chae04}, \cite{Cao2012}, \cite{Chen2012}, 
\cite{Jackson2012}, \cite{campanelli11}, \cite{maniaratra12}, 
\cite{Poitras2012}, and \cite{pan13}.} However their error bars are still
too large to allow for a discrimination between constant and time-varying 
dark energy densities.

Of course, both methods are equivalent, since they make use of the same data,
but each has it own advantages and disadvantages. In particular, it is of
some interest to actually discern the deceleration-acceleration transition
in the $H(z)$ data. While the data does indicate the transition, see Fig.\ 4
of FR, the data points bounce around quite a bit. Given the low reduced $\chi^2$
for the best-fit models (see FR and Table 1 here), all of which show significant 
evidence for a deceleration-acceleration transition, we investigate different data binning
techniques here, to see if binned versions of the $H(z)$ measurements more 
clearly illustrate the presence of a deceleration-acceleration transition.

Motivated by a similar situation in the early days of CMB anisotropy data 
constraints,\footnote{
Compare the analyses of e.g., \cite{ganga97} and \cite{ratra99}, which 
determine constraints on cosmological model parameters from CMB anisotropy 
data, to that of \cite{Podariu2001}, who bin CMB anisotropy measurements to 
determine a smoother observed CMB anisotropy angular power spectrum.} 
in this paper
we use the binning techniques of \cite{Podariu2001} to bin the $H(z)$ data and
so construct a smoother representation of the observed $H(z)$ function. While
\cite{Podariu2001} considered many more data points then we do here
(142 vs.\ 23), they covered a large range in multipole space with 
$\ell_{\rm max} \approx 370 \ell_{\rm min}$
while we consider a significantly smaller range in redshift space with
$z_{\rm max} \approx 15 z_{\rm min}$. The other striking point is that, unlike in
\cite{Podariu2001} for the CMB anisotropy case, we find here that when 
combining individual $H(z)$ measurements the error bars of these measurements are consistent with what is expected for Gaussian errors. This is quiet reassuring. 
We find that the weighted-mean binned $H(z)$ equally well constrain cosmological
parameters, as well as do the unbinned data. More interestingly, for our purpose here, 
the binned weighted-mean data clearly indicate the presence of the deceleration-acceleration
transition.

Our paper is organized as follows. In the next section we summarize
the two techniques we use to bin the $H(z)$ data and compute
binned results for a variety of data points per bin. In Sec.\ \ref{constraints}
we use the binned $H(z)$ data to derive constraints on cosmological 
parameters of the 3 models we consider and show that for the 
weighted-mean binning these constraints 
are very close to those that follow from the unbinned data. We conclude in 
Sec.\ {\ref{summary}}.

\section{Binning the data}
\label{Binning the data}

The 28 individual $H(z)$ measurements bounce around on the $H(z)/(1+z)$ plot,
Fig.\ 4 of FR. To try to get a smoother observed $H(z)/(1+z)$ function we form
bins in redshift and then combine the data points in each bin to give
a single observed value of $z$, $H(z)$, and $\sigma$ for that bin. The measurements
in each bin are combined using two different statistical techniques, weighted 
mean and median statistics. 

Table \ref{table:WA} lists  the weighted mean results. These results were computed using the 
standard formulae  \citep[see, e.g.,][]{Podariu2001}. That is,
\begin{equation}  
\overline{H}(z)=\frac{\sum_{i=1}^{N} H(z_i)/\sigma_{i}^{2}}{\sum_{i=1}^{N}1/\sigma_{i}^{2}},
\end{equation}
where $N$ is the number of data points in the bin under consideration, $\overline{H}(z)$ is the weighted 
mean of the Hubble parameter in that bin, $H(z_i)$ is the value of the Hubble parameter measured
at redshift $z_i$ and $\sigma_i$ is the corresponding uncertainty. Weighted mean redshifts, denoted 
by $\overline{z}$, were similarly computed,
\begin{equation}
\overline{z}=\frac{\sum_{i=1}^{N} z_{i}/\sigma_{i}^{2}}{\sum_{i=1}^{N}1/\sigma_{i}^{2}}.
\end{equation}
The weighted mean standard deviation, denoted by $\overline{\sigma}$, for each bin was found from
\begin{equation}
\overline{\sigma}=\left(\sum_{i=1}^{N} 1/\sigma_i^{2}\right)^{-1/2}.
\end{equation}
The assumptions underlying use of weighted mean statistics are that the measurement errors
are Gaussian, and there are no systematic errors. Hence, one can compute $\chi^2$, the
goodness-of-fit parameter, for each bin,
\begin{equation}
\chi^2=\frac{1}{N-1}\sum_{i=1}^{N}\frac{[H(z_i)-\overline{H}(z)]^2}{\sigma_{i}^2},
\label{eq.chi}
\end{equation}
which has expected value unity and error $1/\sqrt{2(N-1)}$, so we can use this to determine 
the number of standard deviations that $\chi$ deviates from unity for each bin,
\begin{equation}
N_{\overline\sigma}=|\chi-1|\sqrt{2(N-1)}.
\end{equation}
An unaccounted for systematic error, the 
presence of significant correlations between the measurements, and breakdown of the Gaussian error
assumption for each measurement, are the three factors that can make $N_{\overline\sigma}$ 
much greater than unity.

The second technique 
we use to combine measurements in a bin is median statistics, 
as developed in \cite{Gott2001}.\footnote{For other applications of median
statistics see, e.g., \cite{sereno03}, \cite{chen03b}, \cite{Richards2009},
and \cite{Shafieloo11}.} Table\ \ref{table:Med} lists the median 
statistics results. The median is the value for which there is a $50\%$ 
chance of finding a measurement below or above it. 
It is fair to use median statistics to combine the $H(z)$ data of Table 1 of FR
since we assume that all the measurements are independent and there is no over-all systematic 
error in the $H(z)$ data as a whole \citep[individual measurements can have individual systematic errors,
 for discussion see][]{Chen2011a}.
The median will be revealed as a true value 
as the number of measurements grow to infinity, and this technique reduces the effect of outliers 
of a set of measurements on the estimate of a true value. 
If $N$ measurements are considered, the probability of 
finding the true value between values $N_{i}$ and $N_{i+1}$ (where $i=1,2,...N$) is 
\citep{Gott2001},
\begin{equation}
P_{i}=\frac{2^{-N}N!}{i!(N-1)!}
\end{equation}
This process of finding a median value was used for the redshift and the Hubble parameter,
and the Hubble parameter probability distribution was used to determine  $\sigma$ for each bin. 

We would like to have as many measurements as possible in each bin, 
as well as bins that are as narrow as possible in redshift space. 
Obviously, since these requirements are contradictory, compromise 
is necessary. In addition, we require roughly the same number of 
measurements per bin, so as to have approximately similar errors 
on the binned measurements. As indicated in Table\ \ref{table:WA} 
and Table\ \ref{table:Med} we consider four different binnings 
of the 23 lower redshift, $z<1.04$, measurements; the five higher 
redshift measurements are sparsely spread over  too large a 
redshift range to allow for a useful binning.

The last column of Table\ \ref{table:WA} shows that the first 
two binnings, with approximately 3 and 5 measurements per bin, do 
not show any deviation from what is expected from Gaussian 
errors. On the other hand, the last binning, with about 8 
measurements per bin, appears to be not so consistent with 
the assumption of Gaussian errors. This is likely a  consequence 
of the large width in redshift of these bins, so the measurements 
at the low $z$ end and at the high $z$ end of each bin differ too much to
be combined together. Median statistics does not make
use of the error bars of the individual measurements. As a result, 
it is a more conservative technique and when used to combine data in bins
it results in larger error bars. A comparison of the results in Tables\ 
2 and \ 3 clearly illustrates this point. Fortunately the weighted mean
results we have found show that the individual lower redshift data points have 
reasonable error bars and so there is no obvious danger in 
using the more constraining weighted mean results to draw 
physical conclusions.

The weighted-mean and median 
statistics binned results of Tables\ 2 and 3 are plotted in the 
top panels of Figs.\ \ref{fig:For table 2,3}---\ref{fig:For table 8,9}
(in purple).
These figures also show the 5 higher $z$ unbinned measurements listed 
in Table\ 1 of FR (in cyan). Both sets of observations show $1$
and $2$ $\sigma$ error bars. Also shown are the unbinned data (Table\ 1 of FR)
best-fit predictions for $\Lambda$CDM (red), XCDM (blue), and $\phi$CDM (green)
for the two priors, $\overline{H}_0\pm\sigma_{H_0}$= 68 
$\pm$ 2.8 km s$^{-1}$ Mpc$^{-1}$ (dashed lines) and $\overline{H}_0\pm\sigma_{H_0}$= 7.8 
$\pm$ 2.4 km s$^{-1}$ Mpc$^{-1}$ (dotted lines), from FR. Focusing on the 
weighted-mean panels in each of these plots, and comparing to Fig.\ 4 of FR,
we see that the binned data of Figs.\ 1---3 clearly demarcates a declaration-acceleration
transition.

\section{Constraints from the binned data}
\label{constraints}

In this section we use the weighted-mean and median statistics binned data to derive constraints 
on cosmological parameters of $\Lambda$CDM, XCDM, and $\phi$CDM, and compare these constraints
to those that follow from the unbinned data of Table\ 1 of FR.  

In order to derive constraints on the parameters $\textbf{p}$ of the dark energy models 
discussed above, using the binned 
data from Tables\ \ref{table:WA} and \ref{table:Med}, we follow the procedure of \cite{Farooq2012}.
The observational data consist of measurements 
of the Hubble parameter $H_{\rm obs}(z_i)$ at redshifts $z_i$, with the corresponding 
one standard deviation uncertainties $\sigma_i$. To constrain parameters 
of cosmological models, we define the posterior likelihood function 
$\mathcal{L}_{H}(\textbf{p})$, that depends only on the model parameters $\textbf{p}$, 
by integrating the product of the $H_0$ prior likelihood function 
$\propto$ exp$[-(H_0-\bar H_0)^2/(2\sigma^2_{H_0})]$ and the usual likelihood function exp$(-\chi_H^2 /2)$, 
as in Eq.\ (18) of \cite{Farooq2012}. 
Two different Gaussian priors, $\overline{H}_0\pm\sigma_{H_0}$= 68 
$\pm$ 2.8 km s$^{-1}$ Mpc$^{-1}$ 
\citep{Chen2011a} and 
$\overline{H}_0\pm\sigma_{H_0}$ = 73.8 $\pm$ 2.4 km s$^{-1}$ Mpc$^{-1}$
\citep{Riess2011} are used in the marginalization of the likelihood function over the nuisance 
parameter $H_0$. 

The best-fit point (BFP) $\mathbf{p_0}$ are those parameter values that
maximize the likelihood function $\mathcal{L}_H(\mathbf{p})$.
To find the 1, 2, and 3 $\sigma$ confidence intervals as two-dimensional 
parameter sets, we start from the BFP and integrate the volume under $\mathcal{L}_H(\mathbf{p})$  
until we include  68.27, 95.45, and 99.73 \% of the probability.

The lower 3 rows of panels in Figs.\ \ref{fig:For table 2,3}---\ref{fig:For table 8,9} show the 
constraints (1, 2, and 3 $\sigma$ contours)
from the unbinned $H(z)$ data of Table\ 1 of FR (in blue dot-dashed contours) and from the binned $H(z)$ 
data of Tables \ref{table:WA} and \ref{table:Med} here (in red solid contours), for the three dark energy 
models we consider, and for the two different $H_0$ priors mentioned above. 
The red filled circles and the blue empty circles are the best fit points for the binned and 
unbinned data respectively. Some relevant results are listed in Tables\ 4---7.
Comparing the weighted-mean BFP cosmological parameter values listed in these tables, to those listed in the captions of Figs.\ 1---3 of FR, establishes the very good 
agreement between the values derived here using the binned data (especially
for fewer measurements per bin) and the FR values derived
using the unbinned data.

It is clear from the left two columns of the lower three rows of 
Figs.\ \ref{fig:For table 2,3}---\ref{fig:For table 8,9} that the weighted-mean
binning of the first 23 data points in Table\ 1 of FR give almost exactly the 
same constraints on model parameters $\textbf{p}$ for the three cosmological 
models as do the unbinned data of Table\ 1 of FR.
Since the weighted-mean technique
assumes that the error in the measurements has a Gaussian distribution
and that the measurements are uncorrelated, this 
result is consistant with this assumption that the 
$H(z)$ data in Table\ 1 of FR have Gaussian errors. Consequently 
the best way to combine the measurements in a bin is to use the weighted-mean method. 
It is also useful to note that when there are fewer data points in a narrower bin, 
the constraints from the binned data matches better with the constraints derived 
from the unbinned data. This is not unexpected. In the case of median 
statistics, however, the constraints on model parameters for all three models 
from the binned data are much less restrictive than those 
derived from the unbinned data, see the right hand column of panels in the lower three rows of Figs\ \ref{fig:For table 2,3}---\ref{fig:For table 8,9}. This is because median statistics is a more
conservative technique and so, in this case, is not the
best way of combining $H(z)$ measurements in bins. 
It is also interesting to note, from Tables 
\ref{table:fig1 details}---\ref{table:fig4 details}, that $\chi^2_{\rm min}$ for the case of median 
statistics is significantly smaller than $\chi^2_{\rm min}$ for 
the weighted mean case. This is a direct 
consequence of the larger error bars estimated by the more conservative median statistics approach.

\section{Conclusion}
\label{summary}

We have shown that the weighted-mean combinations of the lower redshift $H(z)$
measurements in bins in redshift provide close to identical constraints on 
cosmological model parameters as do the unbinned $H(z)$ data tabulated in FR.
This is consistent with the $H(z)$ measurements errors being Gaussian.

When plotted against $z$, the weighted-mean binned $H(z)/(1+z)$ measurements
bounce around much less than the individual measurements considered in FR, 
and now much more clearly show the presence of a cosmological 
deceleration-acceleration transition, consistent with the new summary 
redshift $z_{\rm da}=0.74 \pm 0.04$ estimated here and consistent with that 
estimated in FR. This result is also consistent with what is
expected in the standard spatially-flat $\Lambda$CDM model and in other cosmological
models with present-epoch energy budget dominated by dark energy.

More, and more precise, measurements of $H(z)$ in the redshift range 
$1\lesssim  z \lesssim 2.5$ will allow for a clearer demarcation of 
the cosmological deceleration-acceleration transition. We anticipate 
that such data will soon become available.

%\acknowledgments

We thank Data Mania, Mikhail Makouski, and Shawn Westmoreland
for useful discussions and helpful advice.
This work was supported in part by DOE grant DEFG03-99EP41093 and 
NSF grant AST-1109275.

%\newpage

%%%%%%%%%%%%%%%
%figure 1
%%%%%%%%%%%%%%%
\begin{figure}[H]
\centering
  \includegraphics[width=62mm]{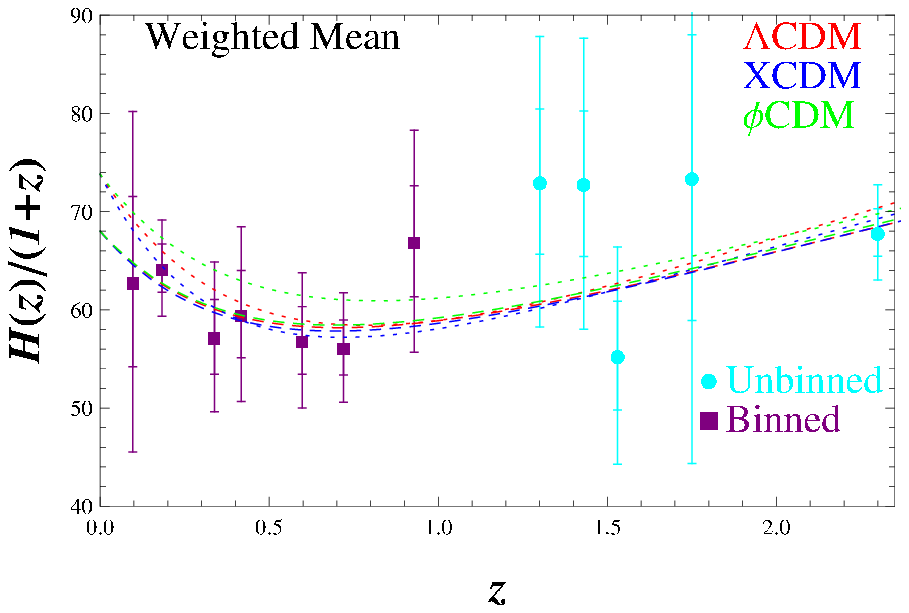}
  \includegraphics[width=62mm]{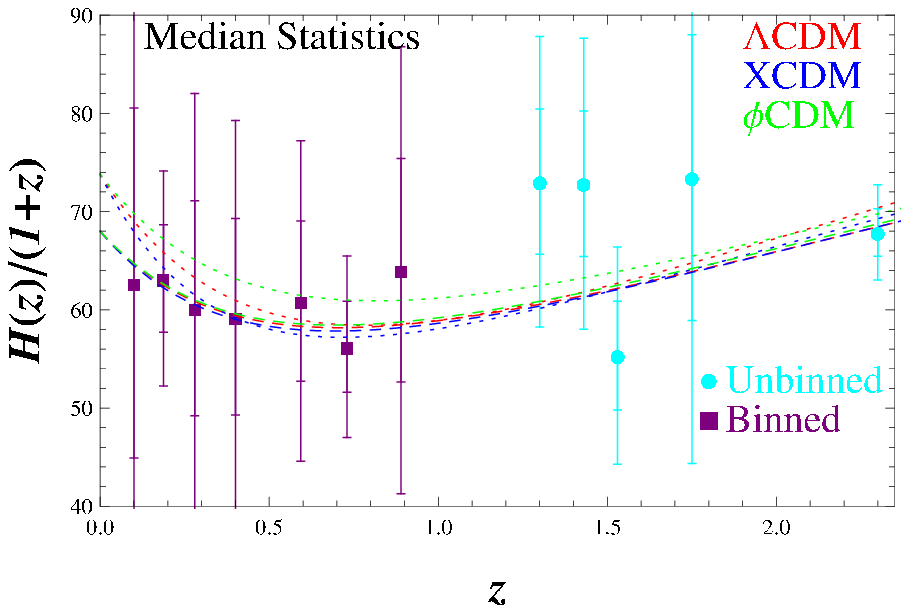}
  \includegraphics[width=39.5mm]{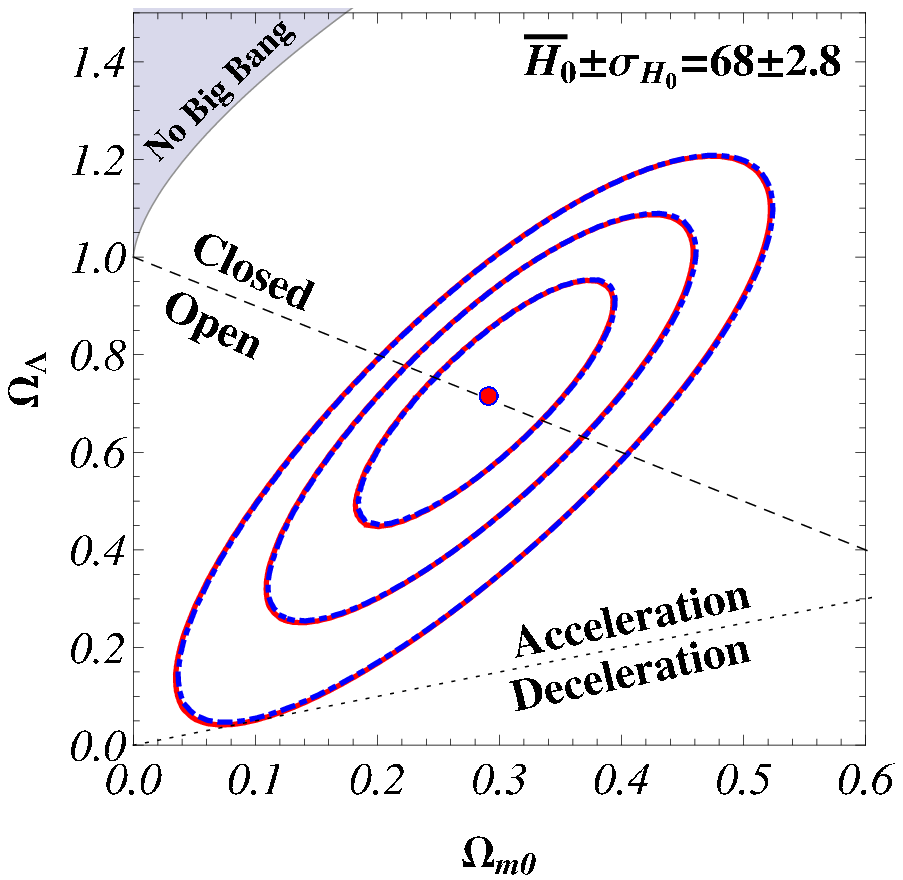}
  \includegraphics[width=39.5mm]{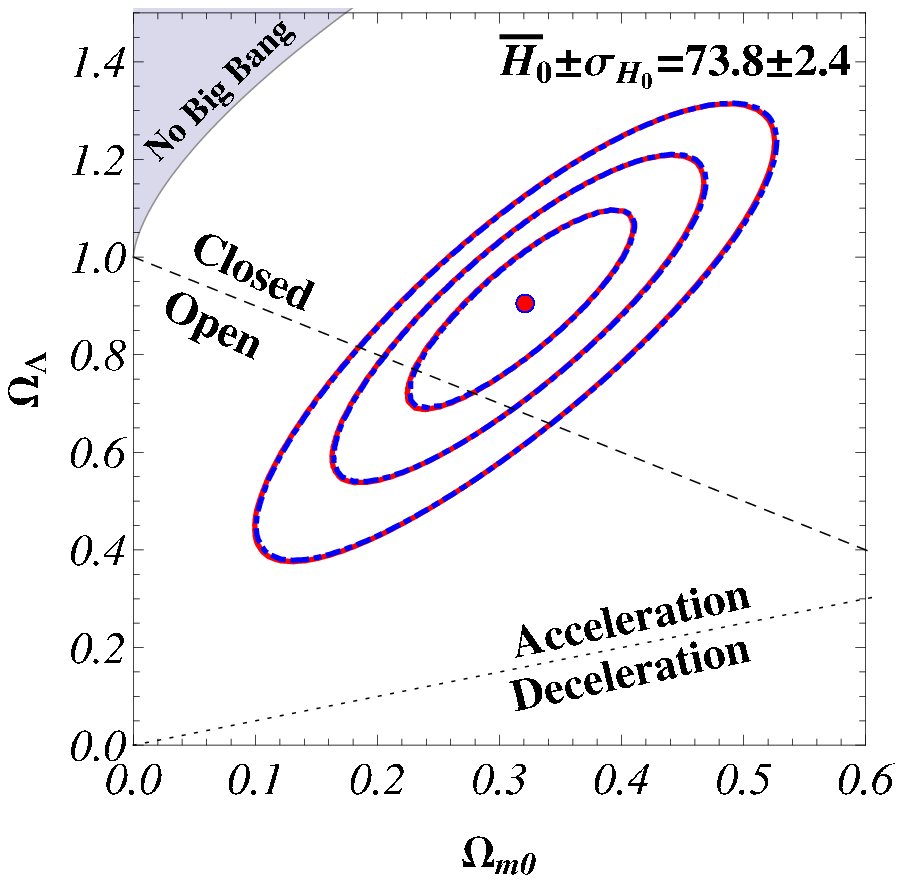}
  \includegraphics[width=39.5mm]{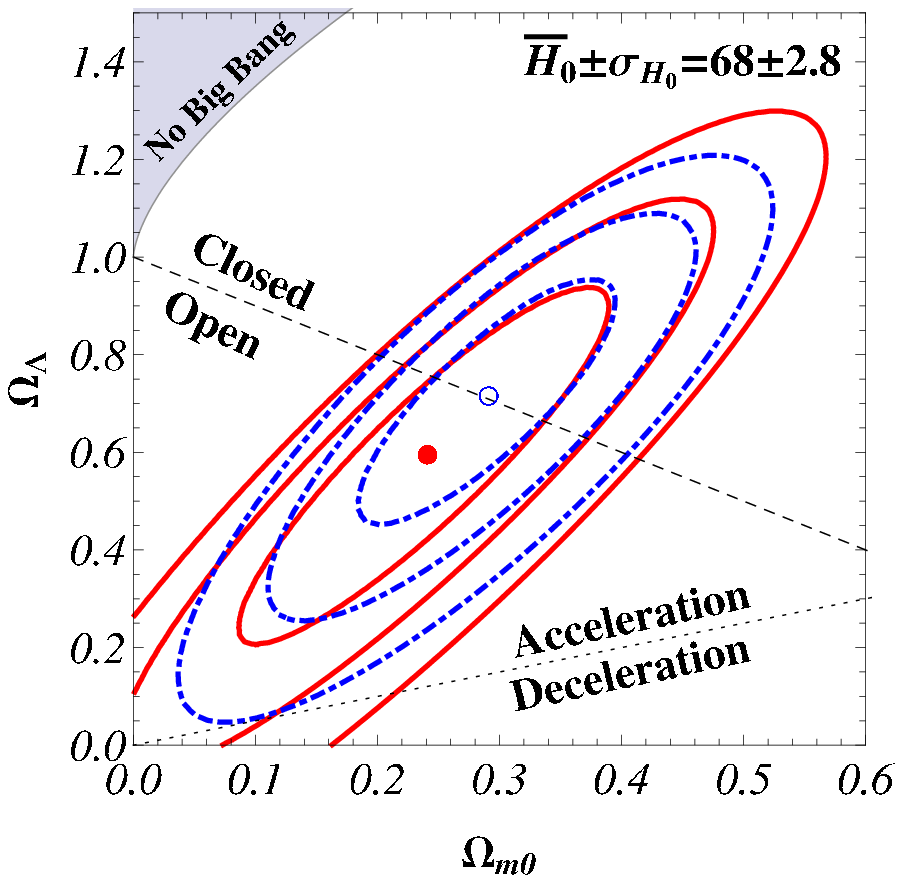}
  \includegraphics[width=39.5mm]{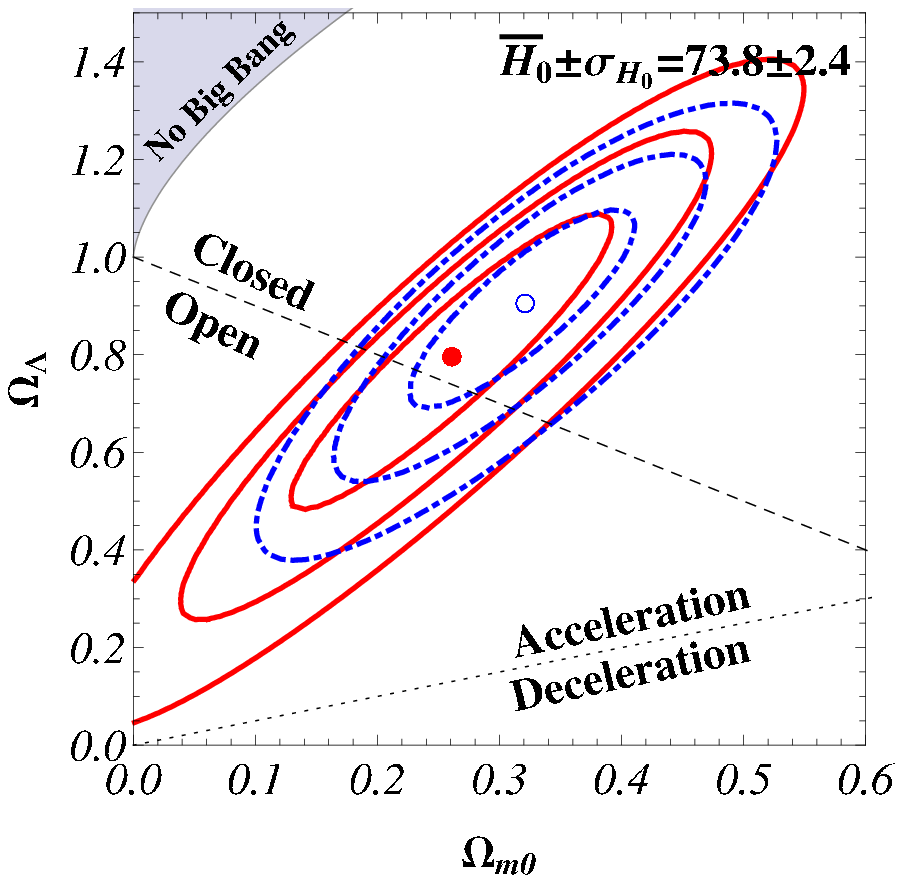}
  \includegraphics[width=40mm]{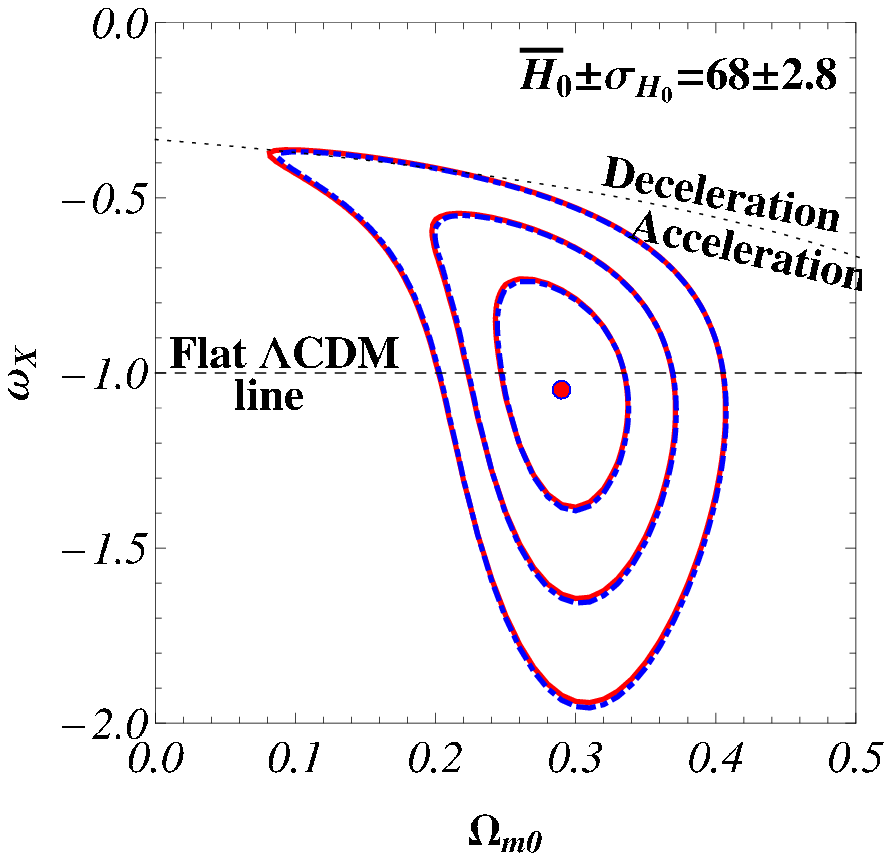}
  \includegraphics[width=40mm]{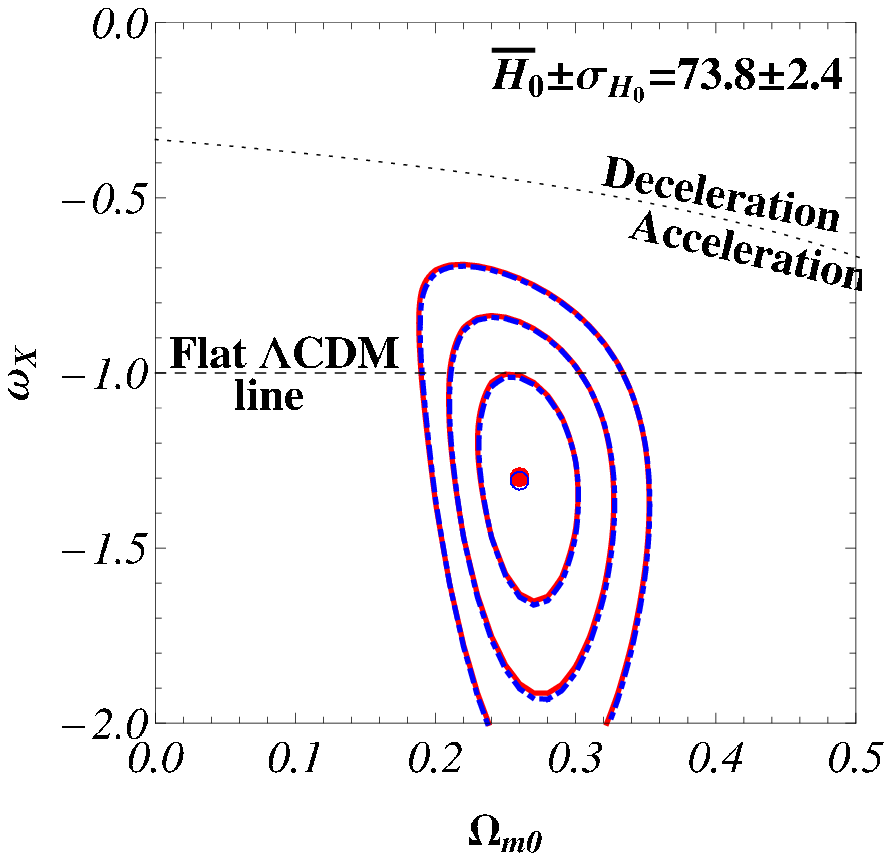}
  \includegraphics[width=40mm]{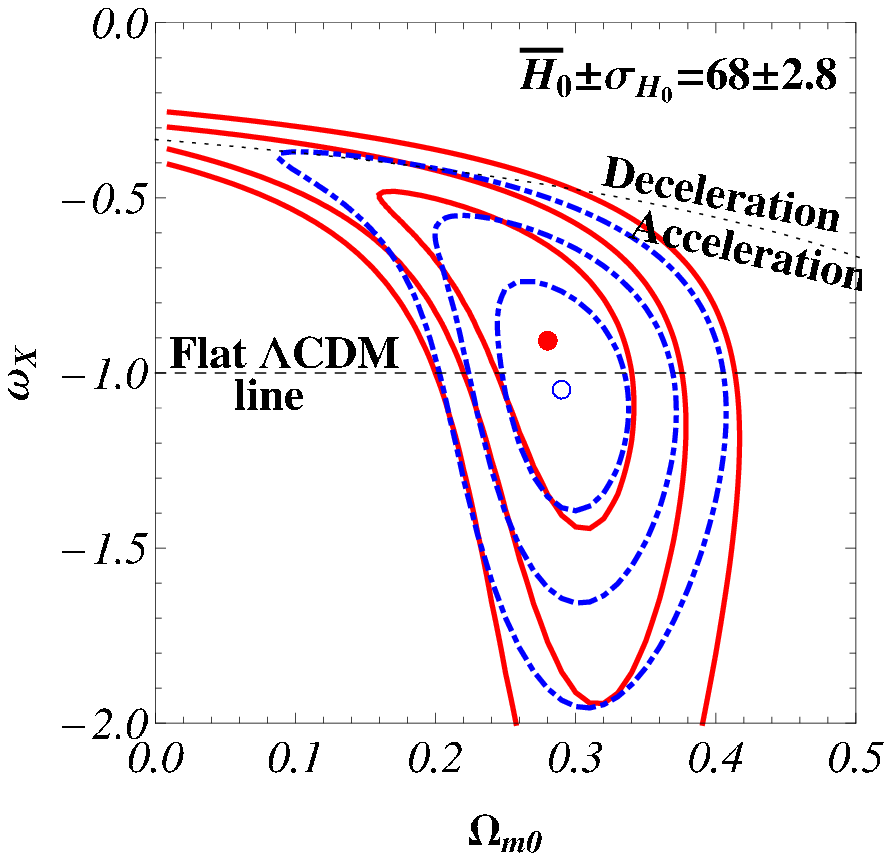}
  \includegraphics[width=40mm]{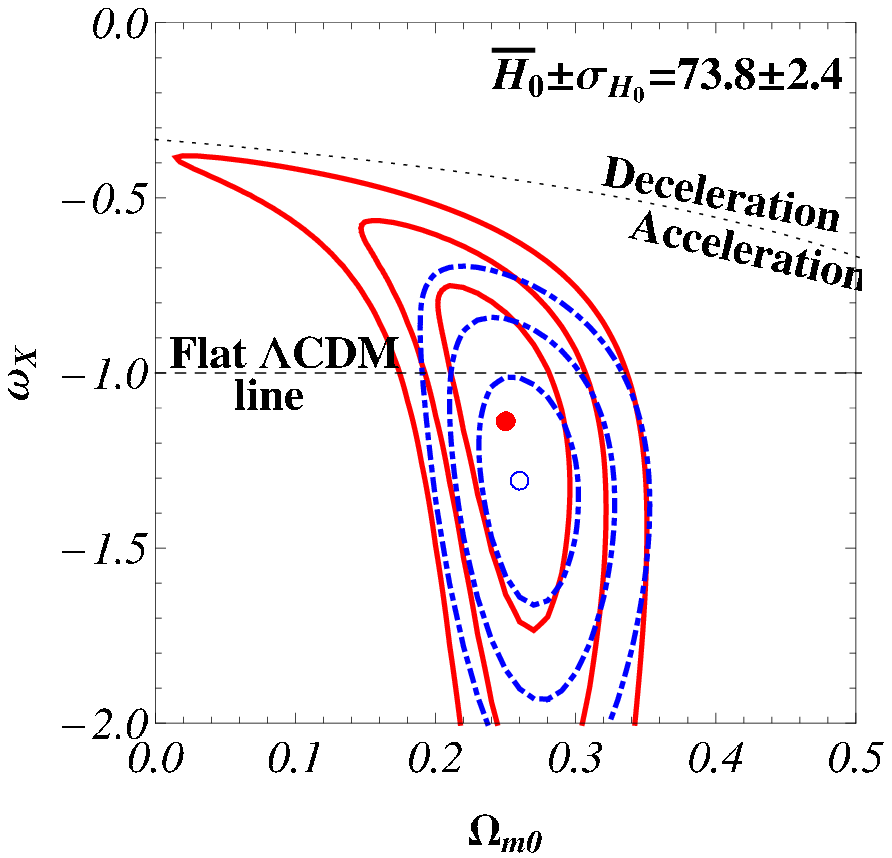}
  \includegraphics[width=39.5mm]{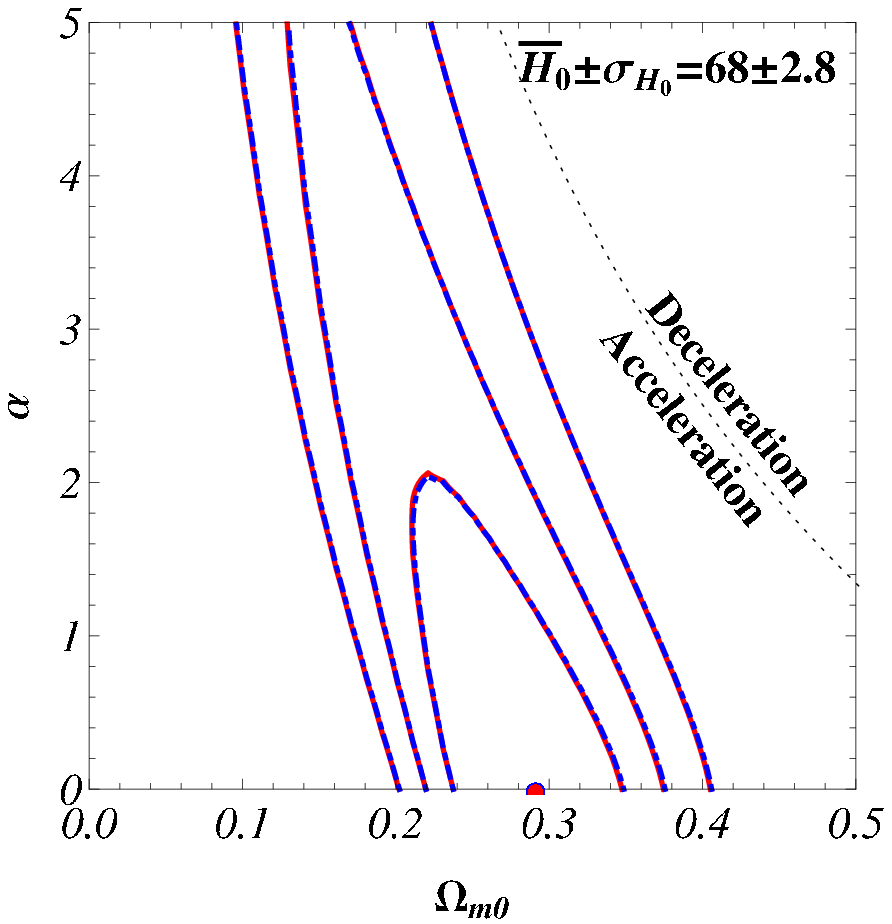}
  \includegraphics[width=39.5mm]{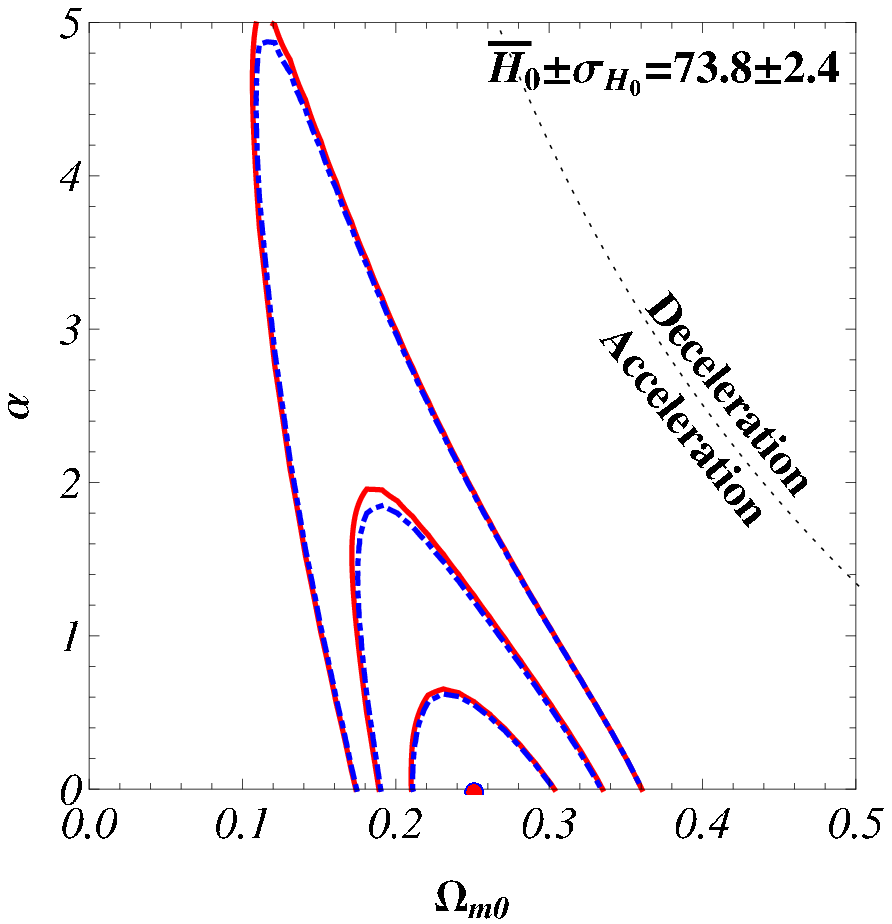}
  \includegraphics[width=39.5mm]{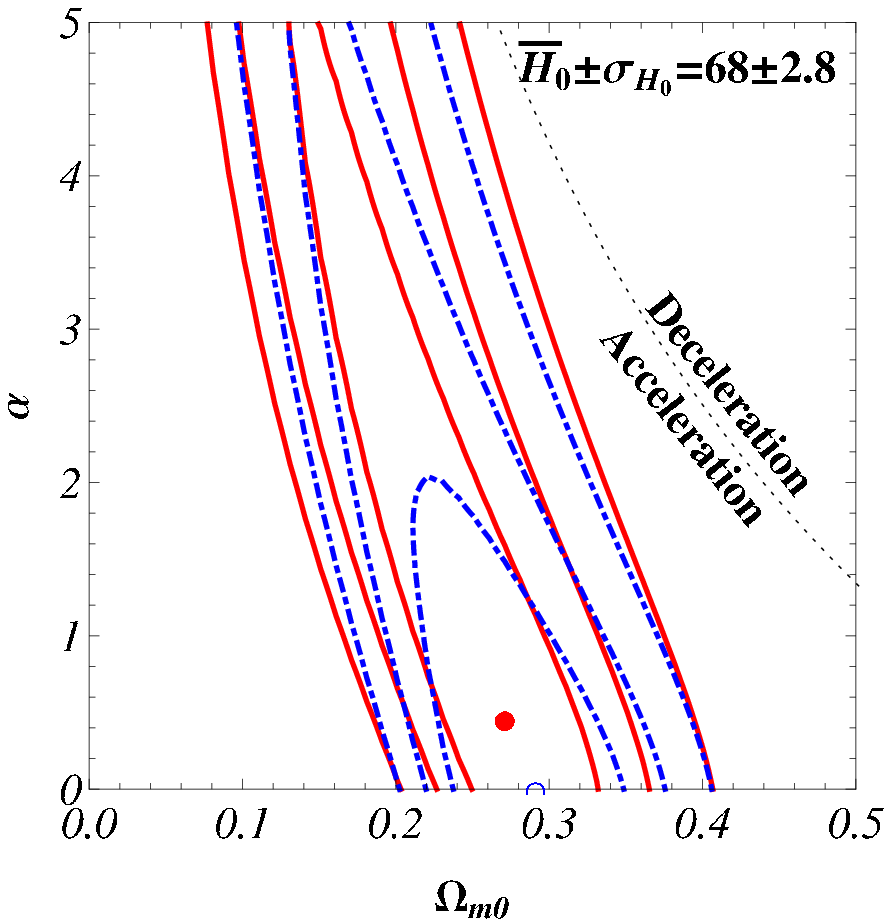}
  \includegraphics[width=39.5mm]{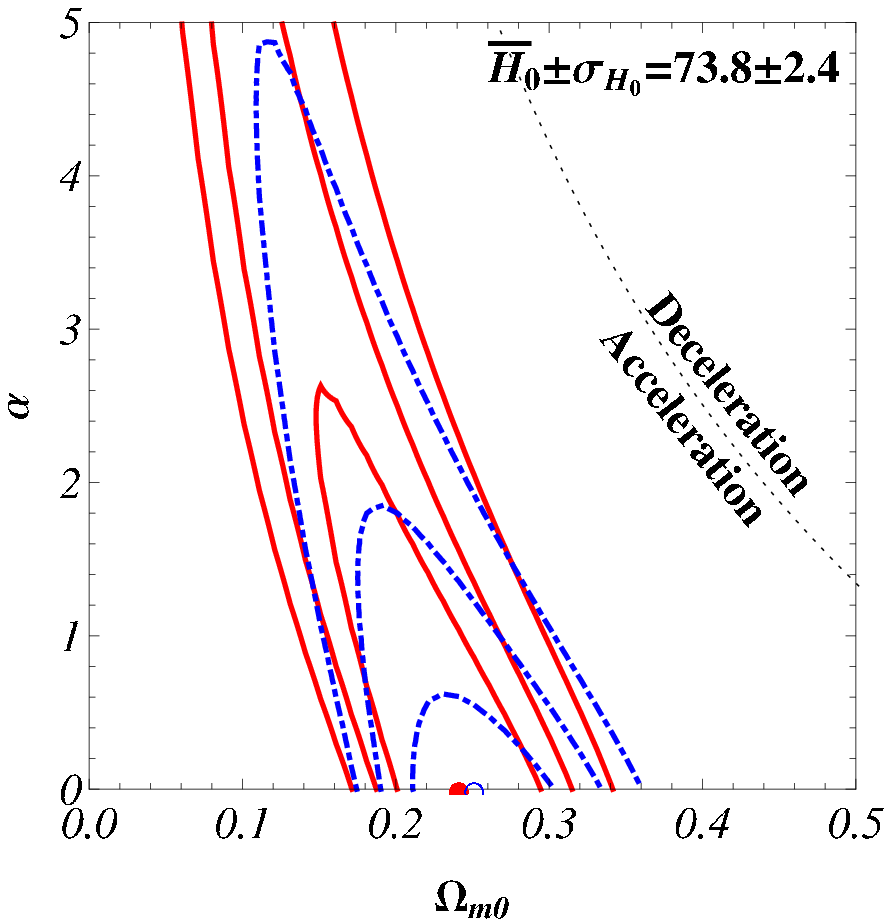}
\caption{
Top left (right) panel shows the $H(z)/(1+z)$ data, binned with 3 or 4 
measurements per bin,  as well as 5 higher $z$ measurements, and the FR 
best-fit model predictions, dashed (dotted) for lower (higher) $H_0$ prior. 
The 2nd through 4th rows show the $H(z)$ constraints for $\Lambda$CDM, 
XCDM, and $\phi$CDM.
Red (blue dot-dashed) contours are 1, 2, and 3 $\sigma$ confidence interval results from 3 or 4
measurements per bin (unbinned FR Table\ 1) data. 
In these three rows, the first two plots 
include red weighted-mean constraints while the second two include red median statistics ones. The 
filled red (empty blue) circle is the corresponding best-fit point.
Dashed diagonal lines show spatially-flat models, and dotted
lines indicate zero-acceleration models. For quantitative details see Table \ref{table:fig1 details}.
} \label{fig:For table 2,3}
\end{figure}

%%%%%%%%%%%%%%%%%%%%
%figure 2
%%%%%%%%%%%%%%%%%%%%
\begin{figure}[H]
\centering
  \includegraphics[width=62mm]{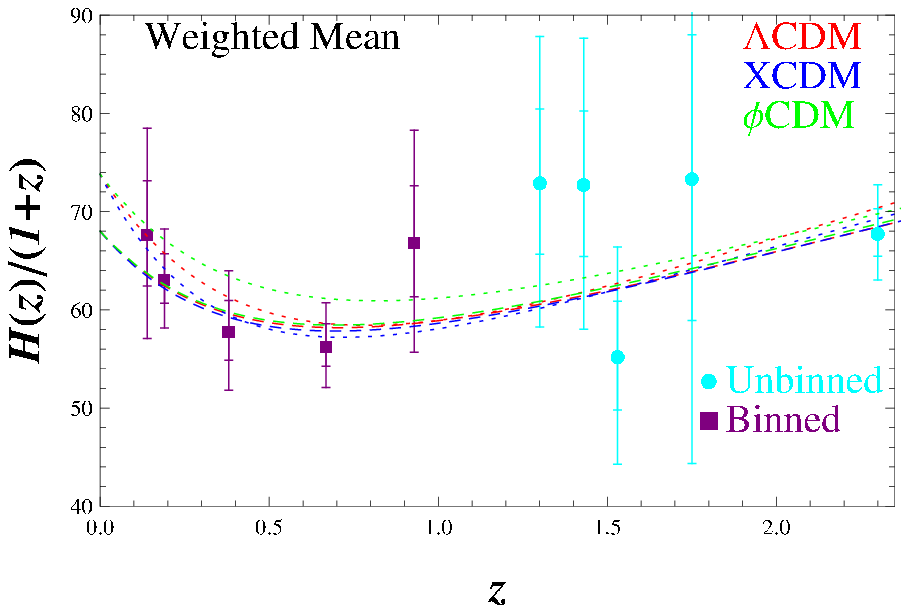}
  \includegraphics[width=62mm]{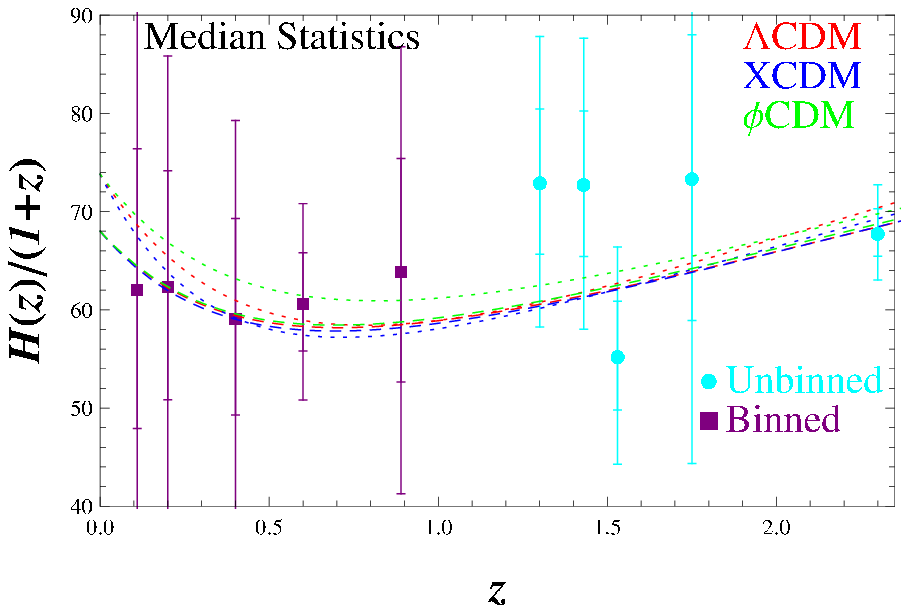}
  \includegraphics[width=39.5mm]{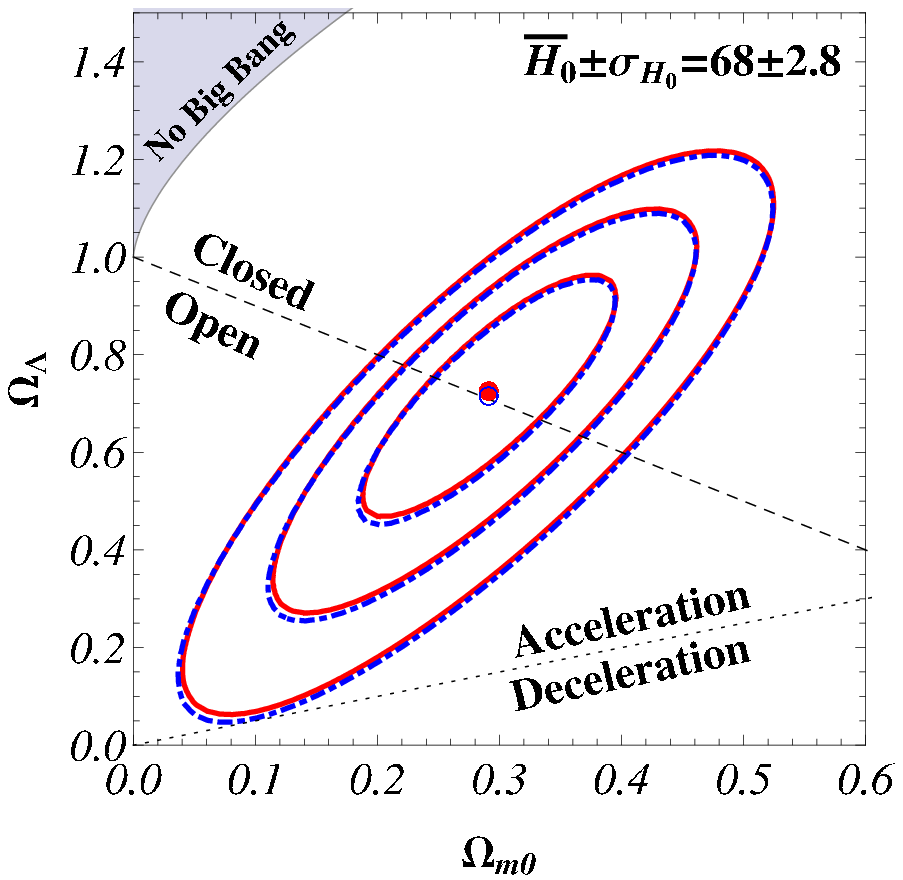}
  \includegraphics[width=39.5mm]{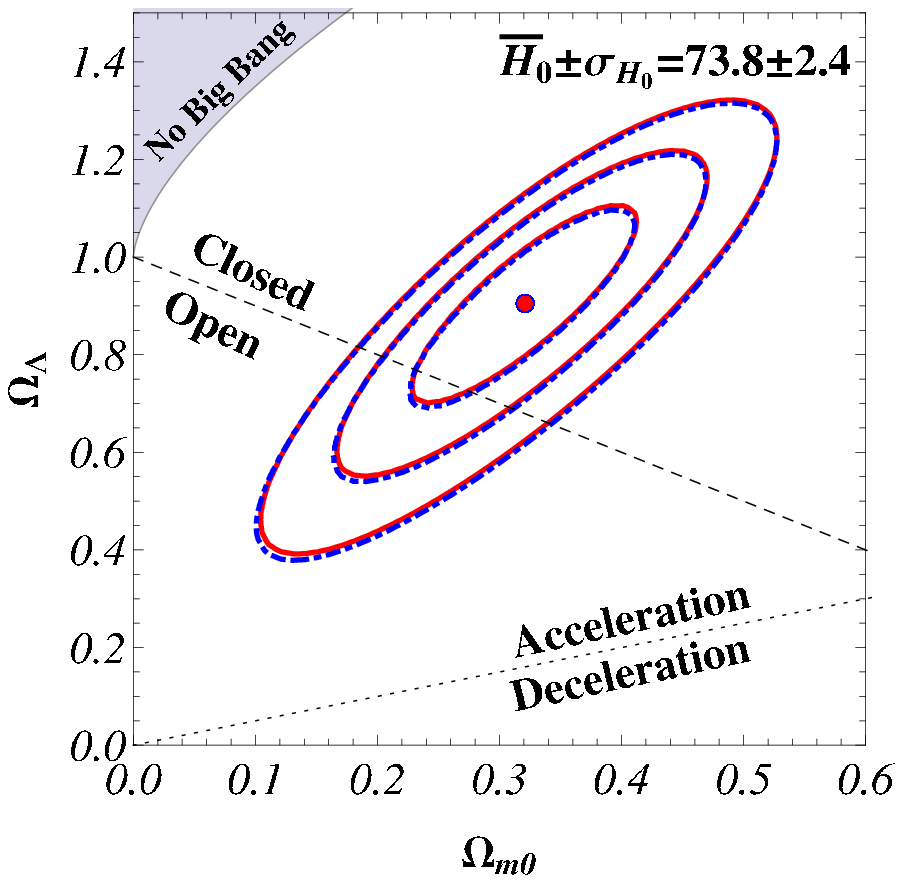}
  \includegraphics[width=39.5mm]{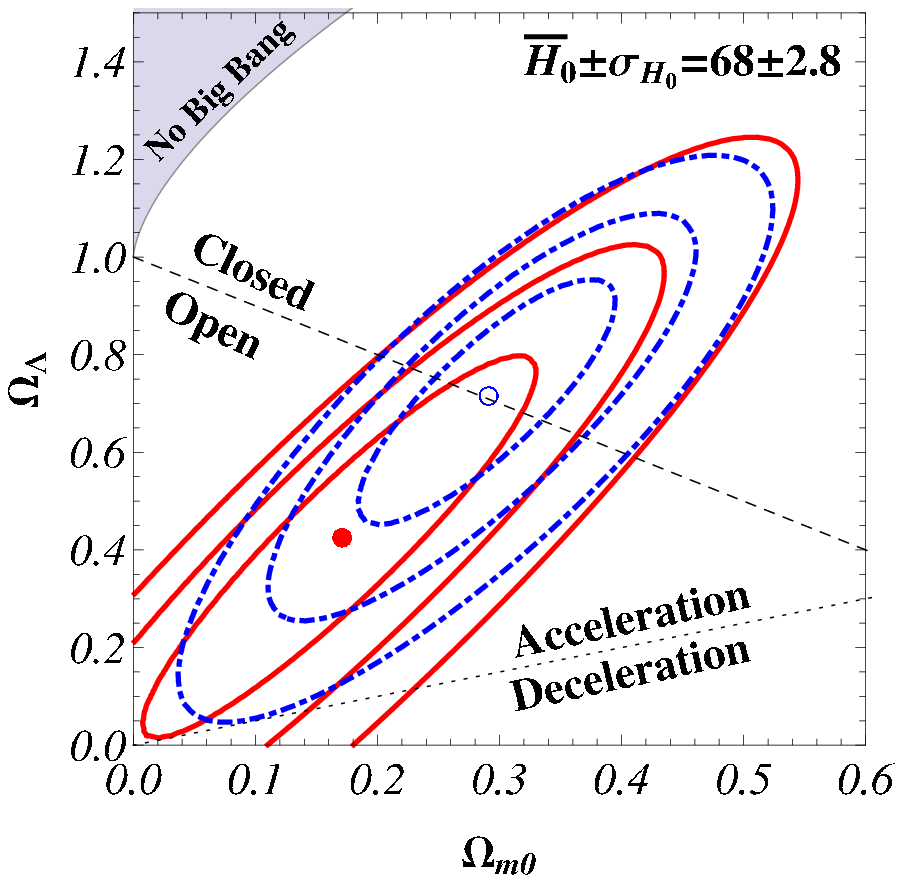}
  \includegraphics[width=39.5mm]{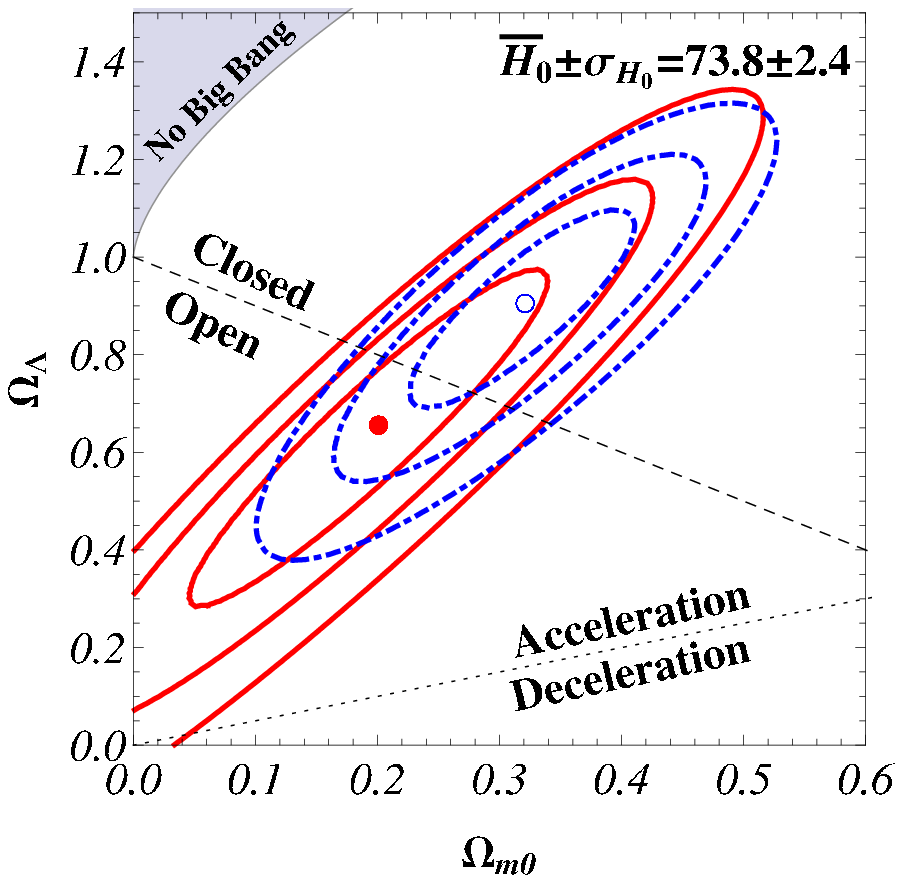}
  \includegraphics[width=40mm]{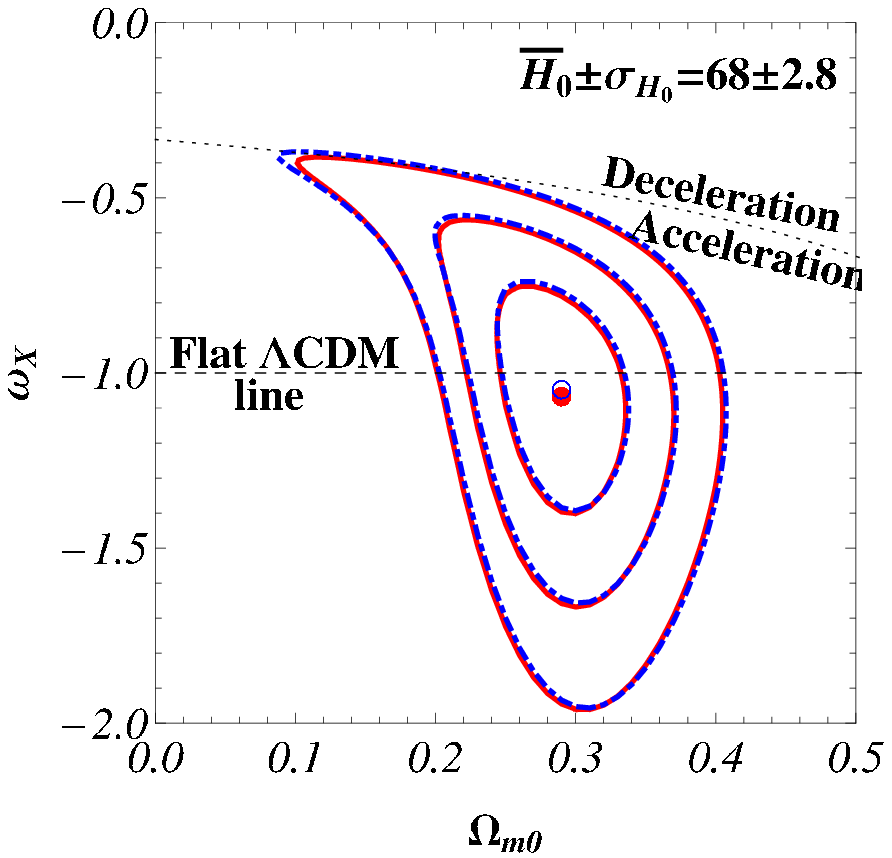}
  \includegraphics[width=40mm]{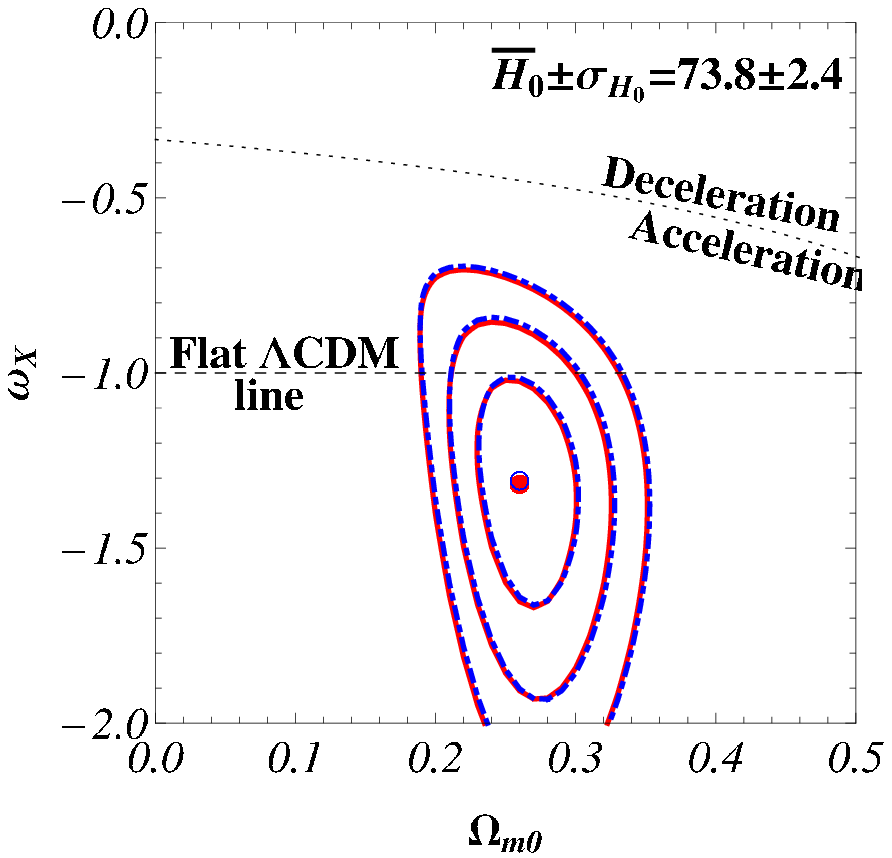}
  \includegraphics[width=40mm]{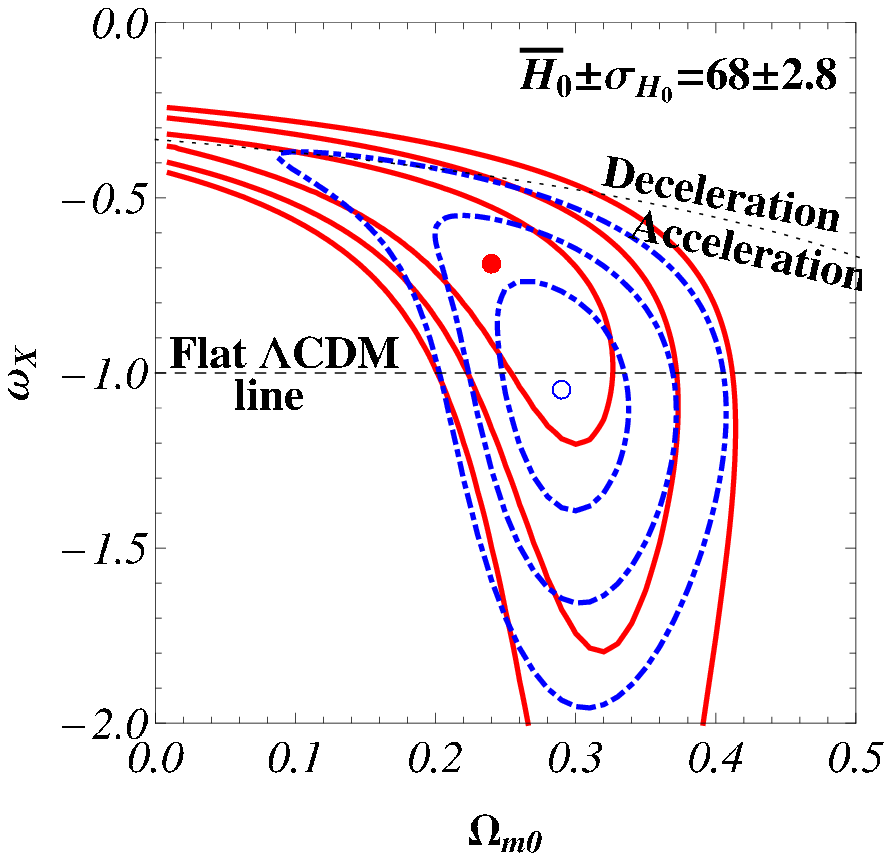}
  \includegraphics[width=40mm]{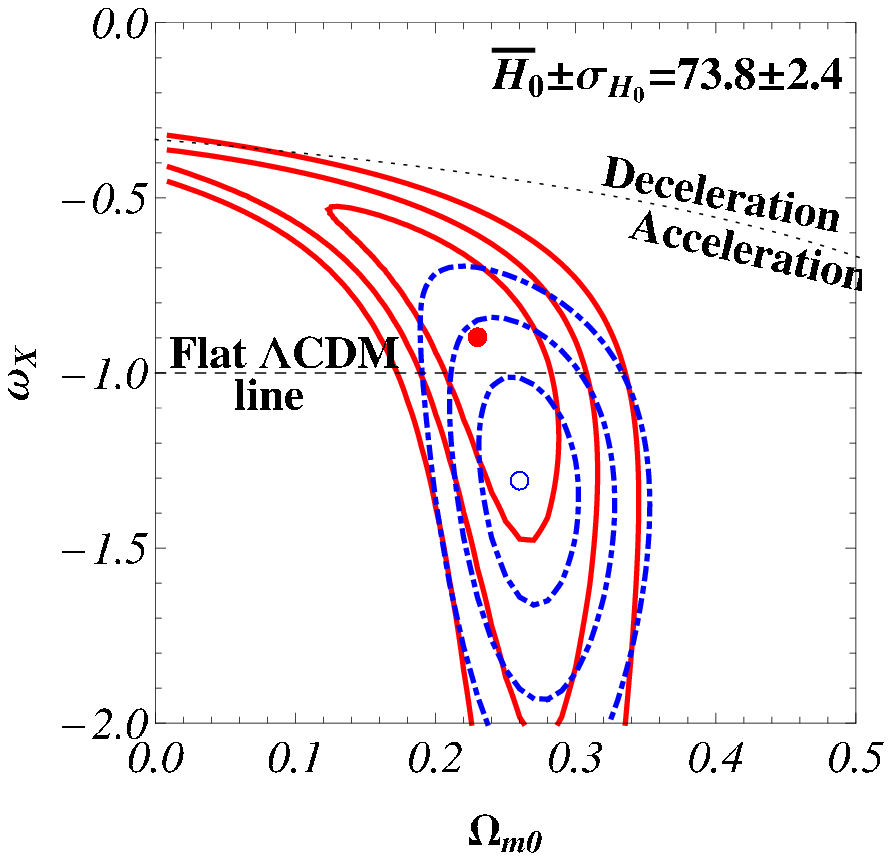}
  \includegraphics[width=39.5mm]{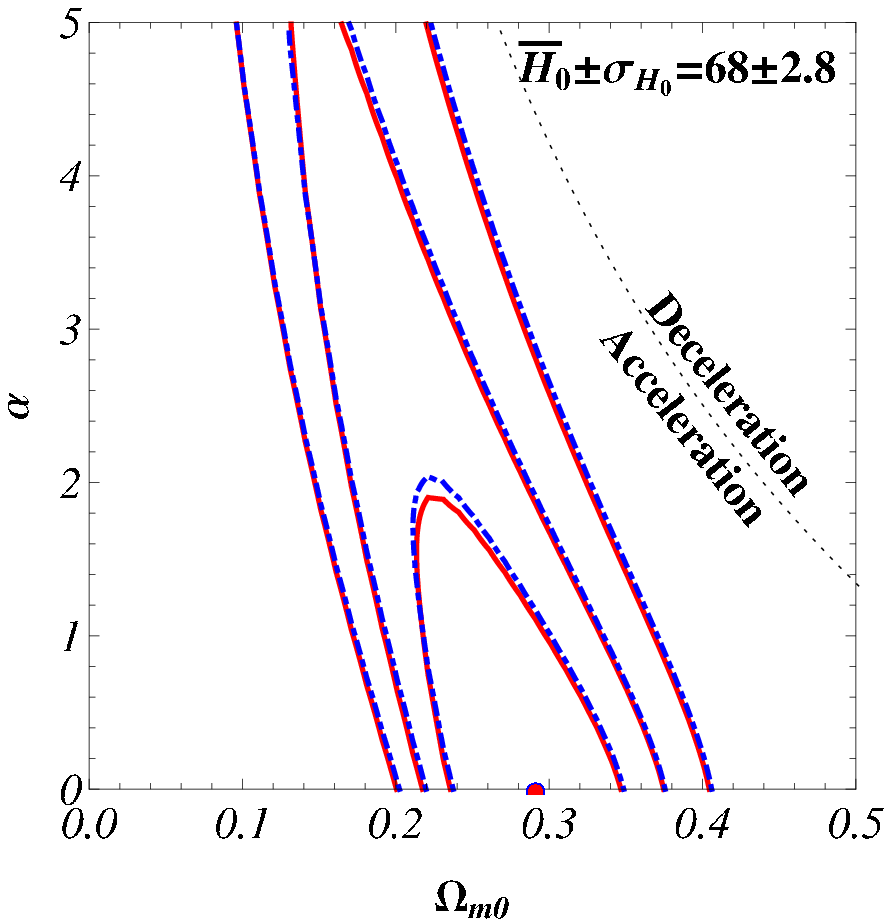}
  \includegraphics[width=39.5mm]{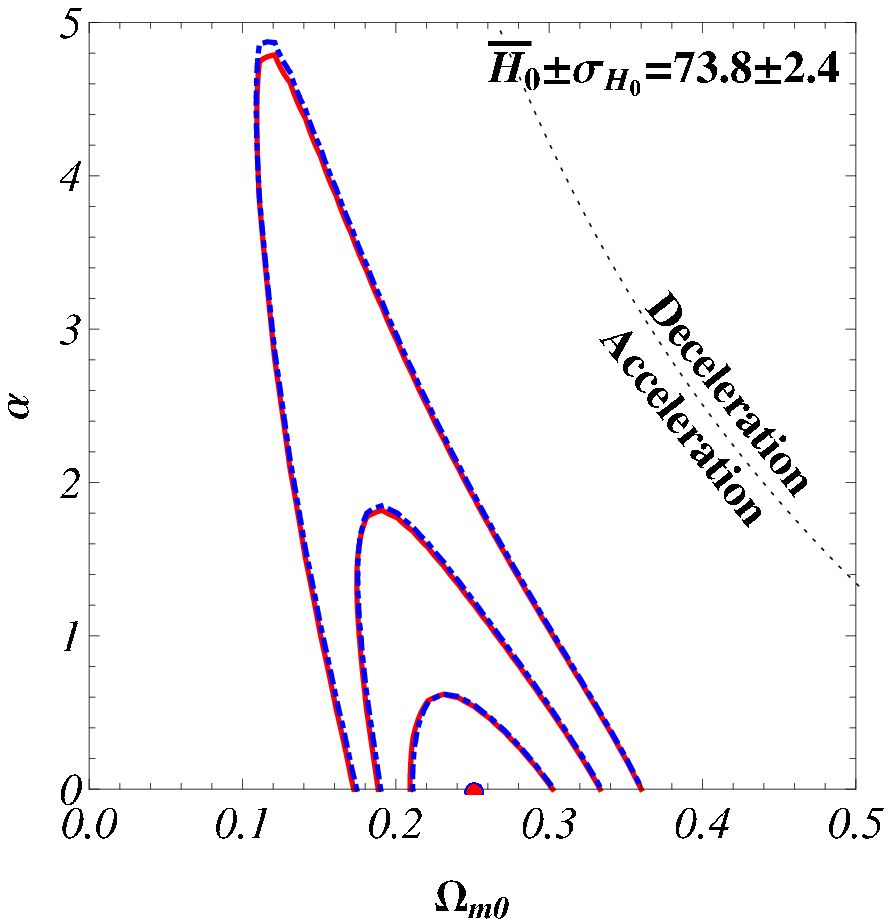}
  \includegraphics[width=39.5mm]{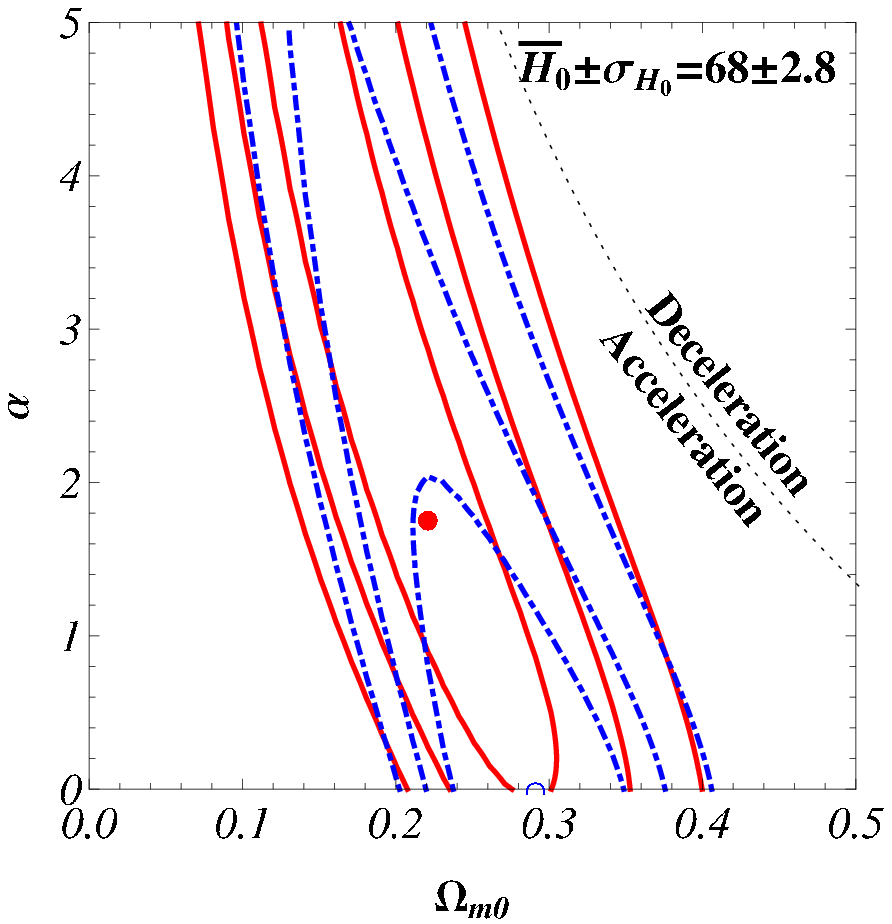}
  \includegraphics[width=39.5mm]{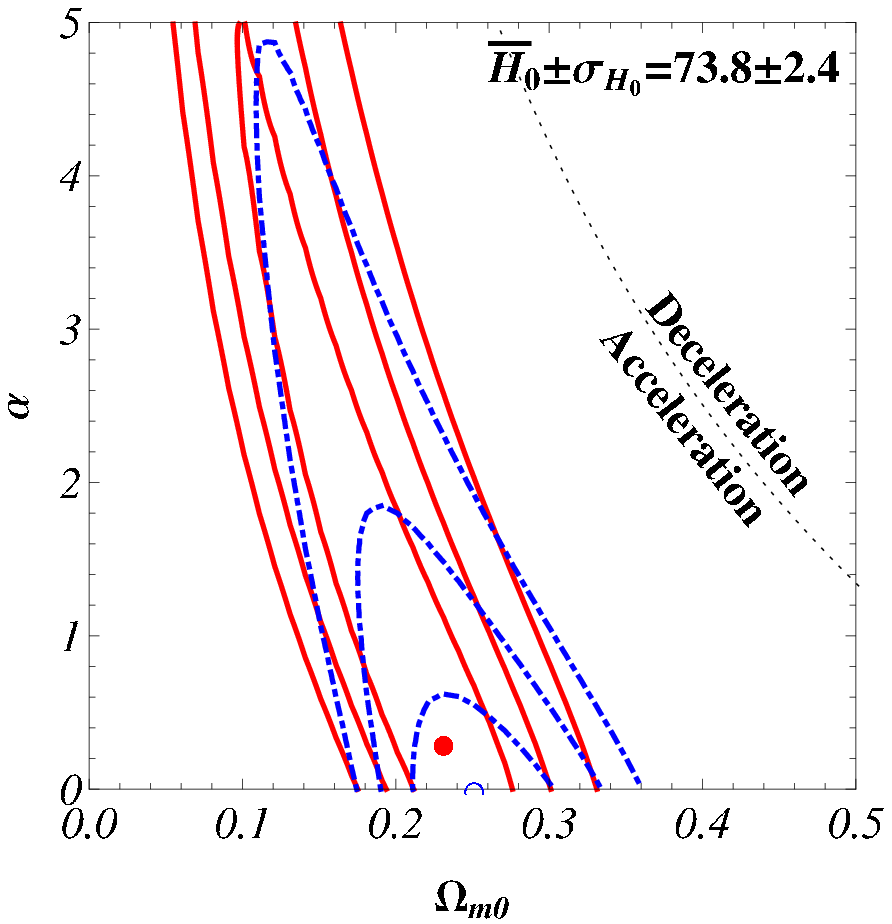}
\caption{
Top left (right) panel shows the $H(z)/(1+z)$ data, binned with 4 or 5 
measurements per bin,  as well as 5 higher $z$ measurements, and the FR 
best-fit model predictions, dashed (dotted) for lower (higher) $H_0$ prior. 
The 2nd through 4th rows show the $H(z)$ constraints for $\Lambda$CDM, 
XCDM, and $\phi$CDM.
Red (blue dot-dashed) contours are 1, 2, and 3 $\sigma$ confidence interval results from 4 or 5
measurements per bin (unbinned FR Table\ 1) data. 
In these three rows, the first two plots 
include red weighted-mean constraints while the second two include red median statistics ones. The 
filled red (empty blue) circle is the corresponding best-fit point.
Dashed diagonal lines show spatially-flat models, and dotted
lines indicate zero-acceleration models. For quantitative details see Table \ref{table:fig2 details}.
} \label{fig:For table 4,5}
\end{figure}

%%%%%%%%%%%%%%%%%%%%%%%%
%figure 3
%%%%%%%%%%%%%%%%%%%%%
\begin{figure}[H]
\centering
  \includegraphics[width=62mm]{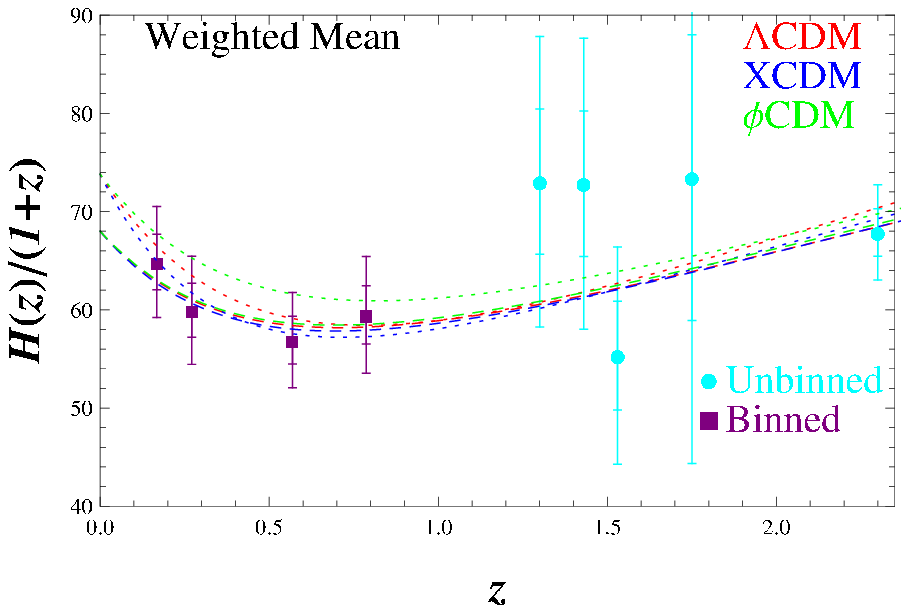}
  \includegraphics[width=62mm]{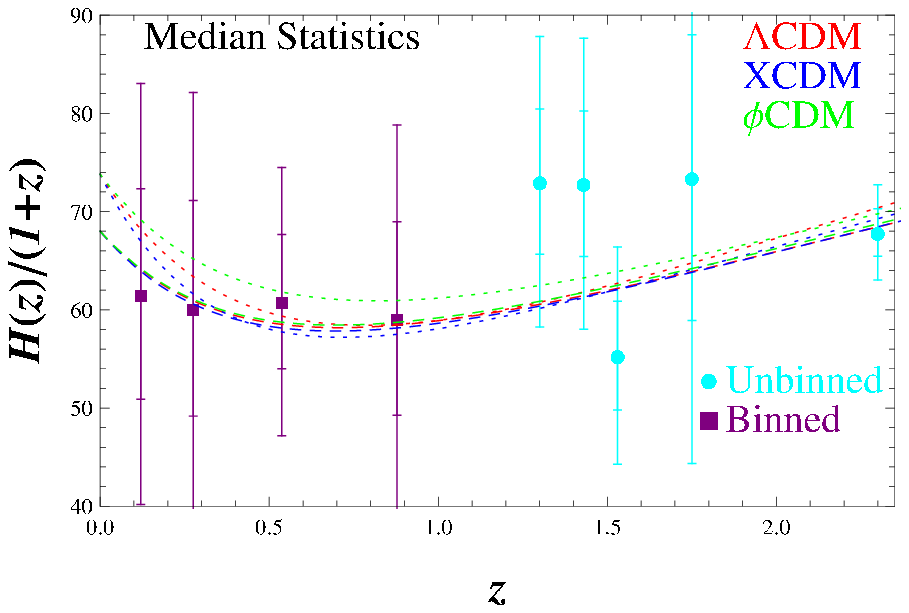}
  \includegraphics[width=39.5mm]{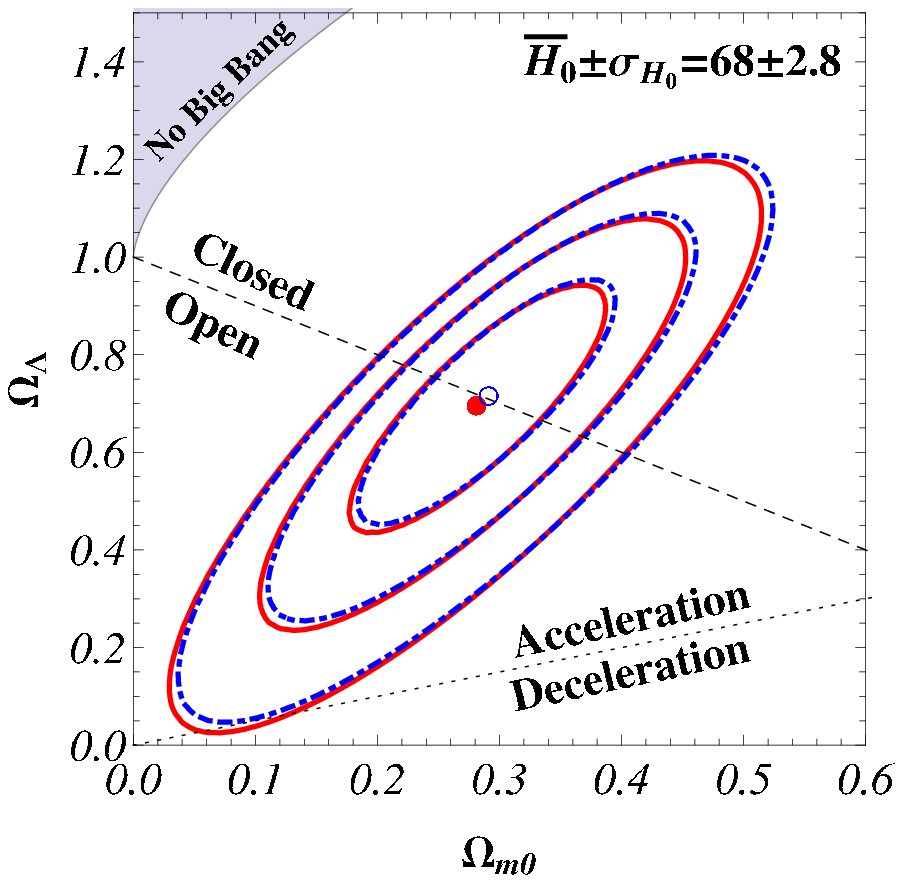}
  \includegraphics[width=39.5mm]{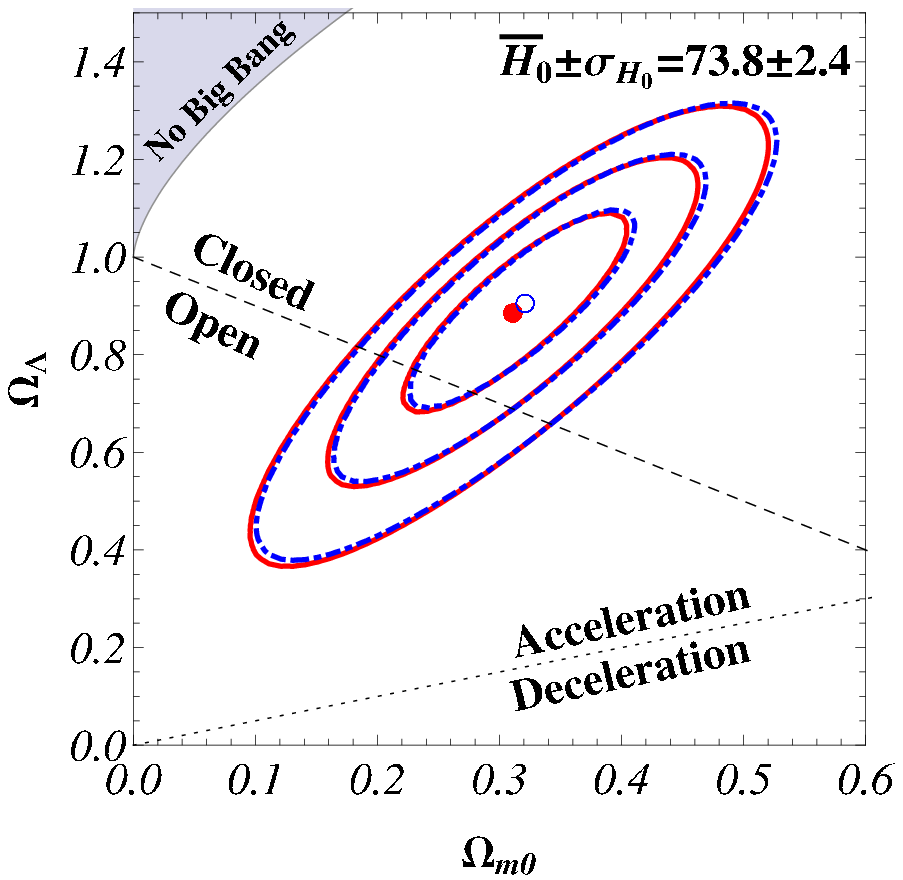}
  \includegraphics[width=39.5mm]{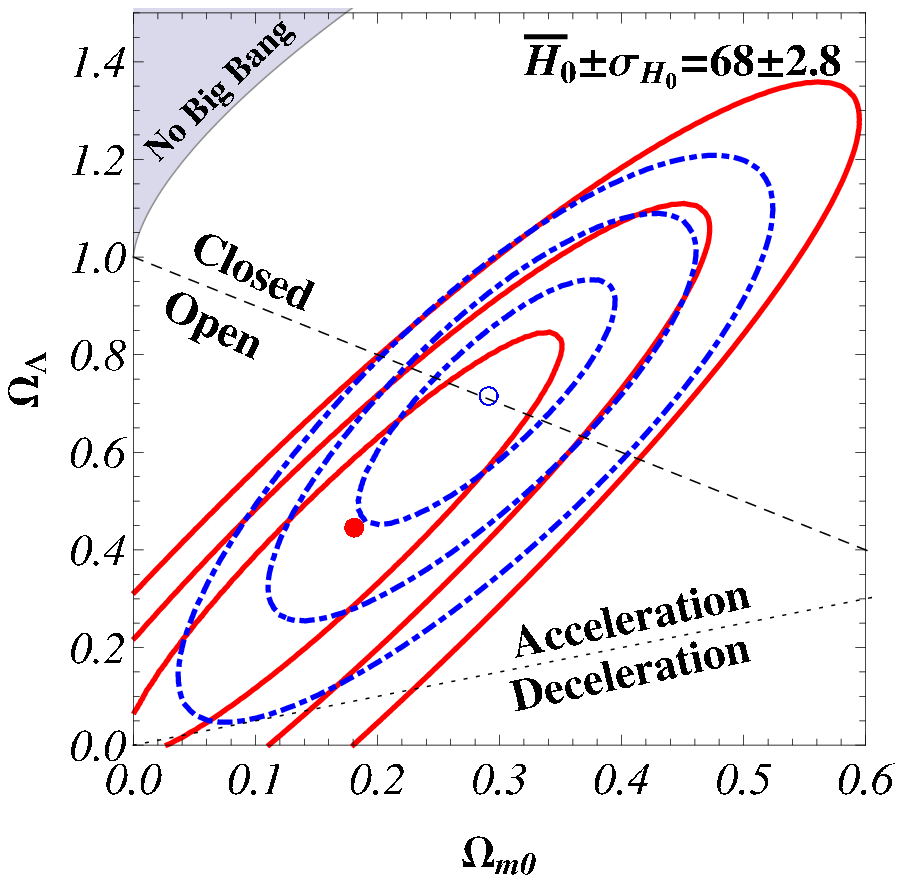}
  \includegraphics[width=39.5mm]{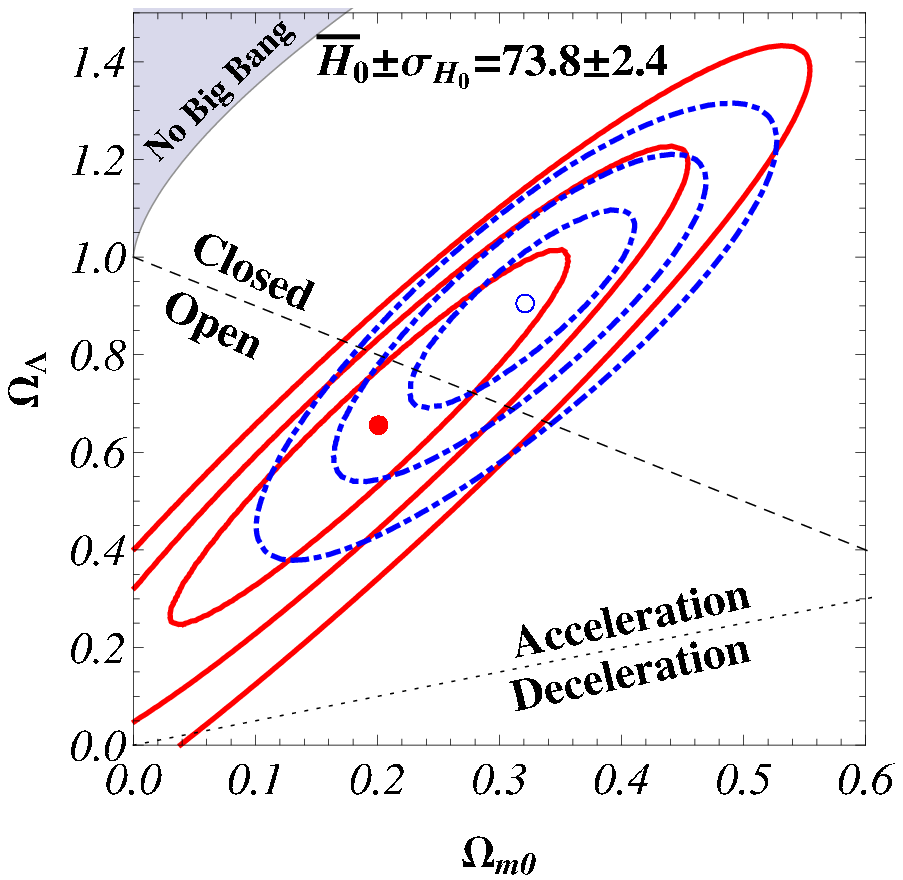}
  \includegraphics[width=40mm]{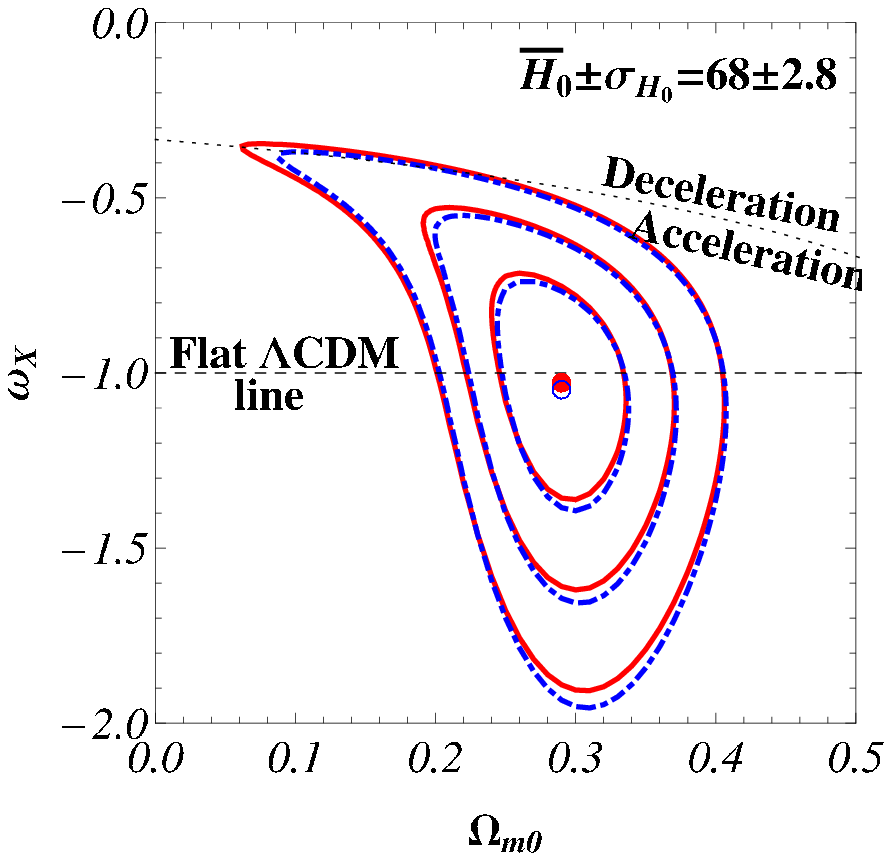}
  \includegraphics[width=40mm]{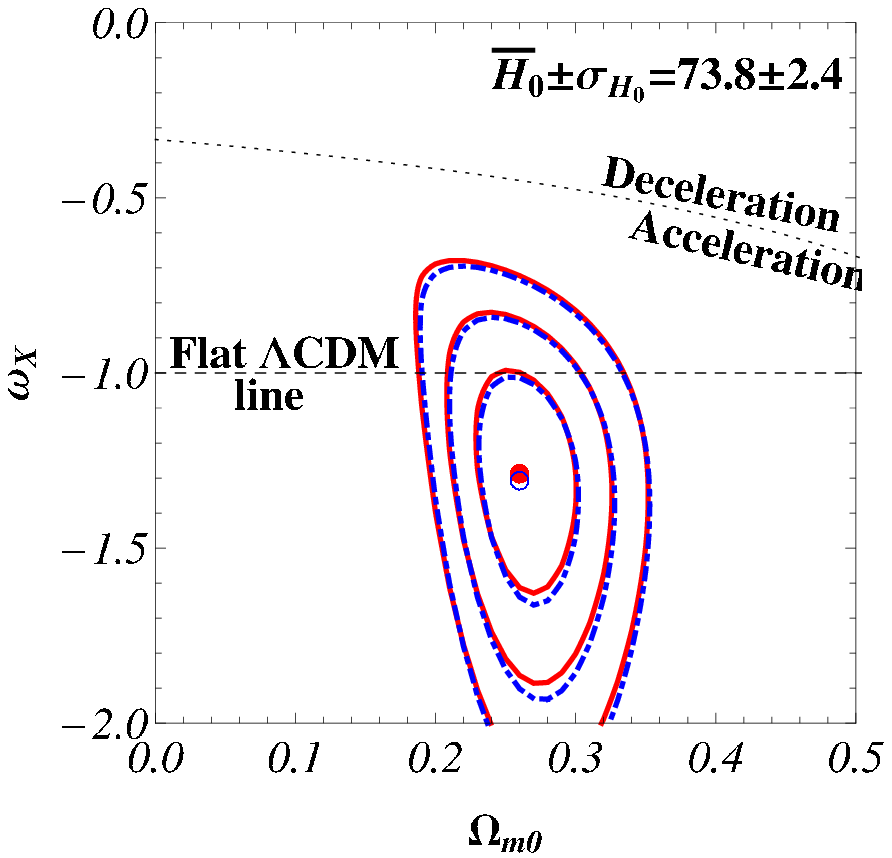}
  \includegraphics[width=40mm]{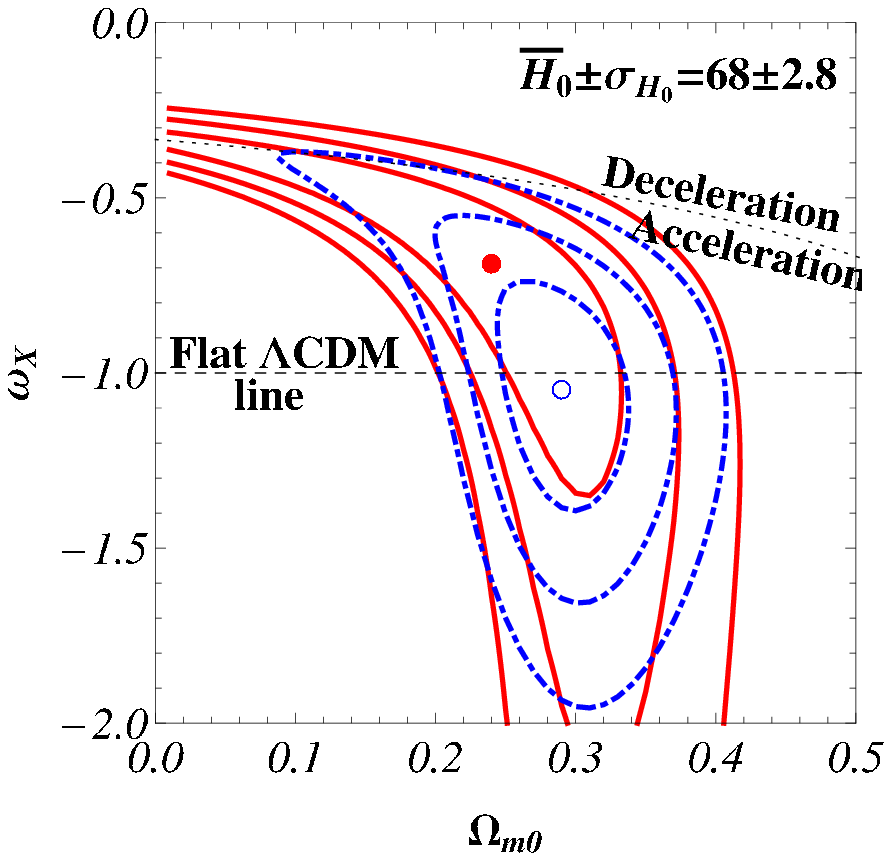}
  \includegraphics[width=40mm]{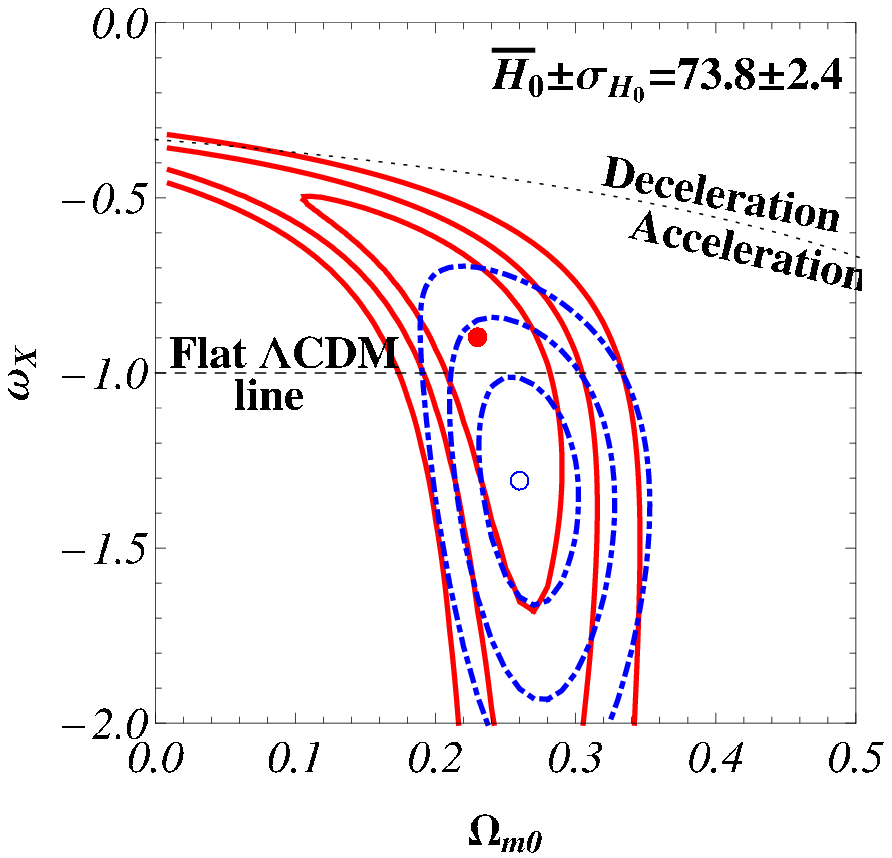}
  \includegraphics[width=39.5mm]{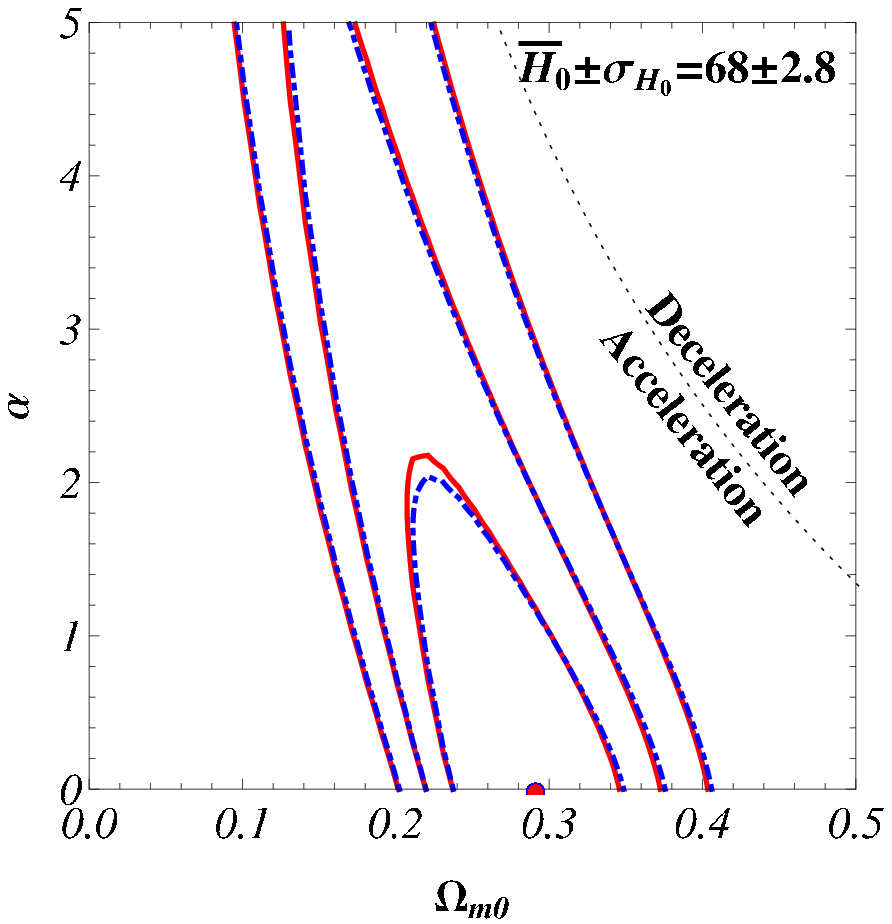}
  \includegraphics[width=39.5mm]{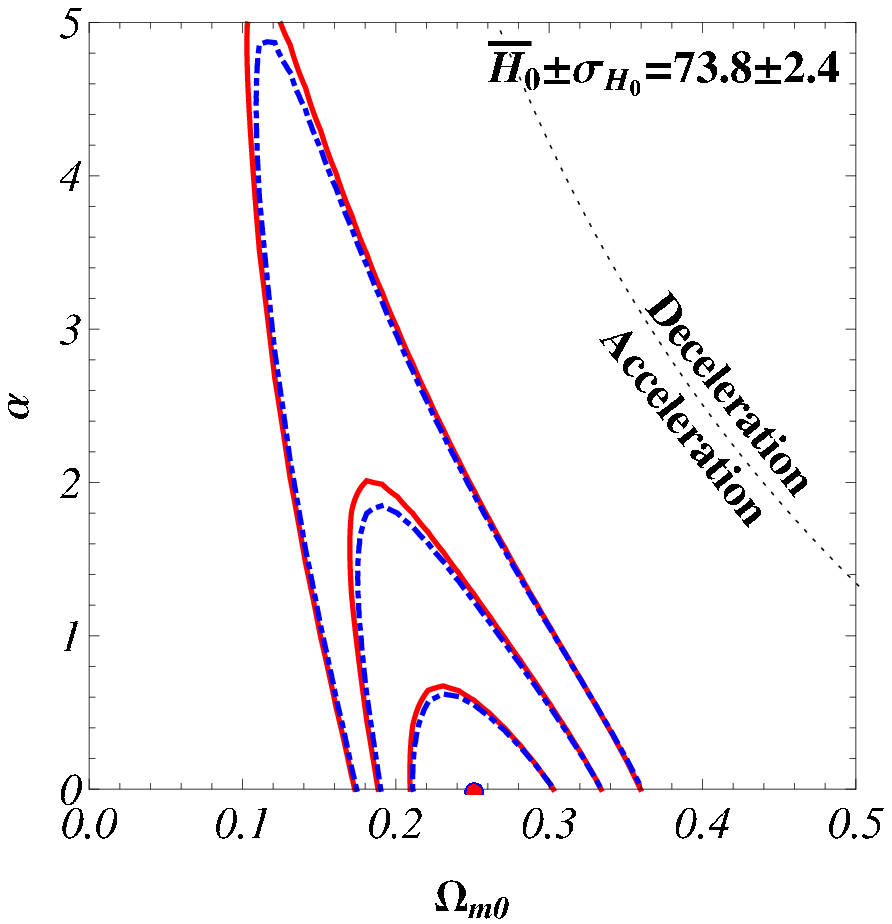}
  \includegraphics[width=39.5mm]{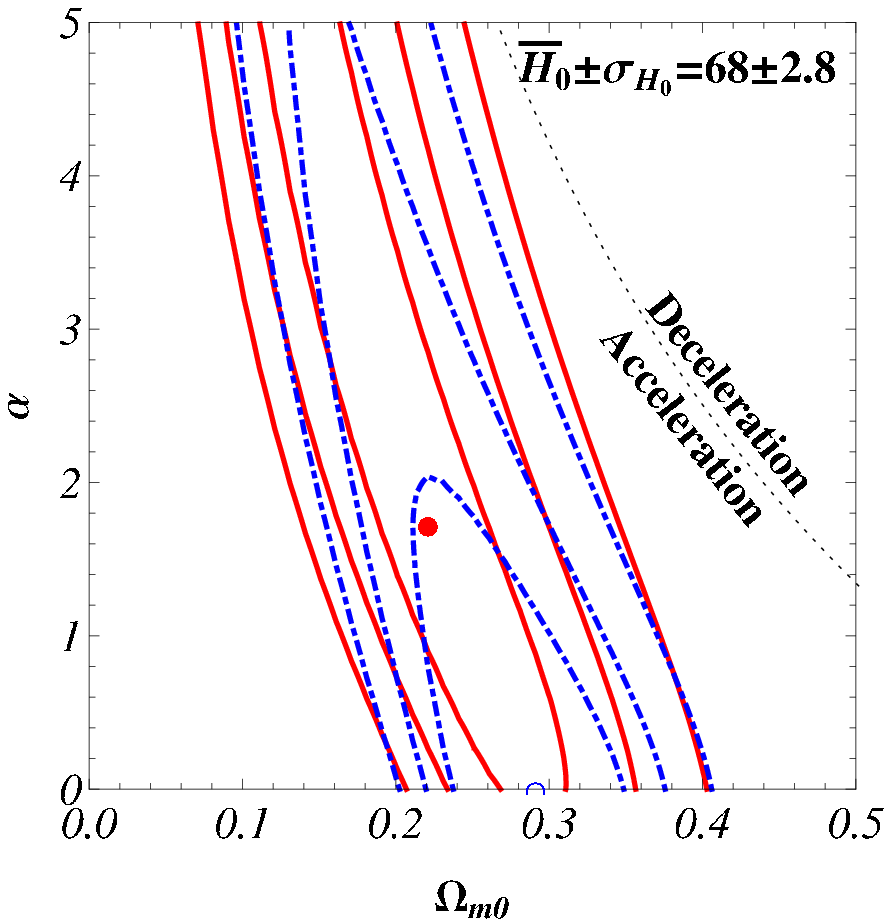}
  \includegraphics[width=39.5mm]{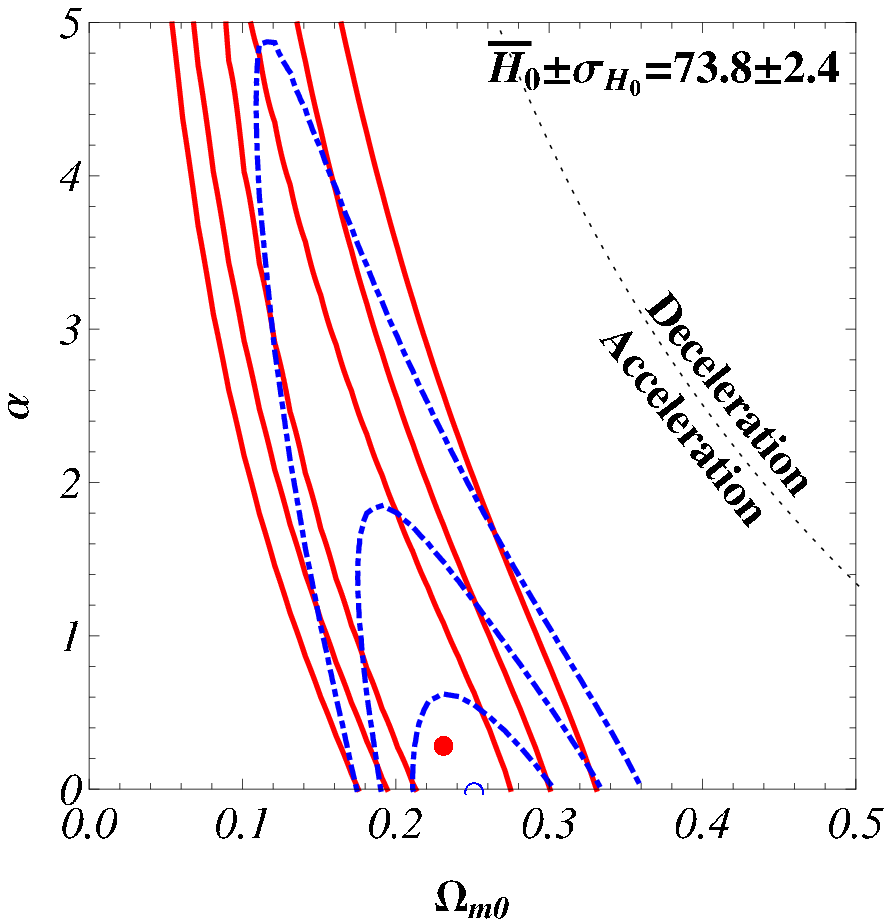}
\caption{
Top left (right) panel shows the $H(z)/(1+z)$ data, binned with 5 or 6 
measurements per bin,  as well as 5 higher $z$ measurements, and the FR 
best-fit model predictions, dashed (dotted) for lower (higher) $H_0$ prior. 
The 2nd through 4th rows show the $H(z)$ constraints for $\Lambda$CDM, 
XCDM, and $\phi$CDM.
Red (blue dot-dashed) contours are 1, 2, and 3 $\sigma$ confidence interval results from 5 or 6
measurements per bin (unbinned FR Table\ 1) data. 
In these three rows, the first two plots 
include red weighted-mean constraints while the second two include red median statistics ones. The 
filled red (empty blue) circle is the corresponding best-fit point.
Dashed diagonal lines show spatially-flat models, and dotted
lines indicate zero-acceleration models. For quantitative details see Table \ref{table:fig3 details}.
} \label{fig:For table 6,7}
\end{figure}

%%%%%%%%%%%%%%%%%%%%%%%%%
%figure 4
%%%%%%%%%%%%%%%%%%%%%
\begin{figure}[H]
\centering
  \includegraphics[width=62mm]{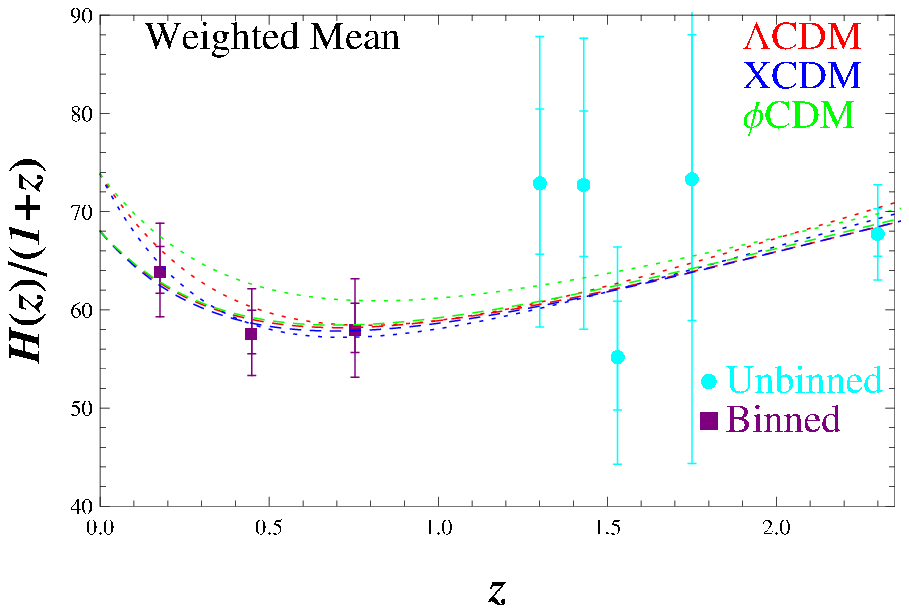}
  \includegraphics[width=62mm]{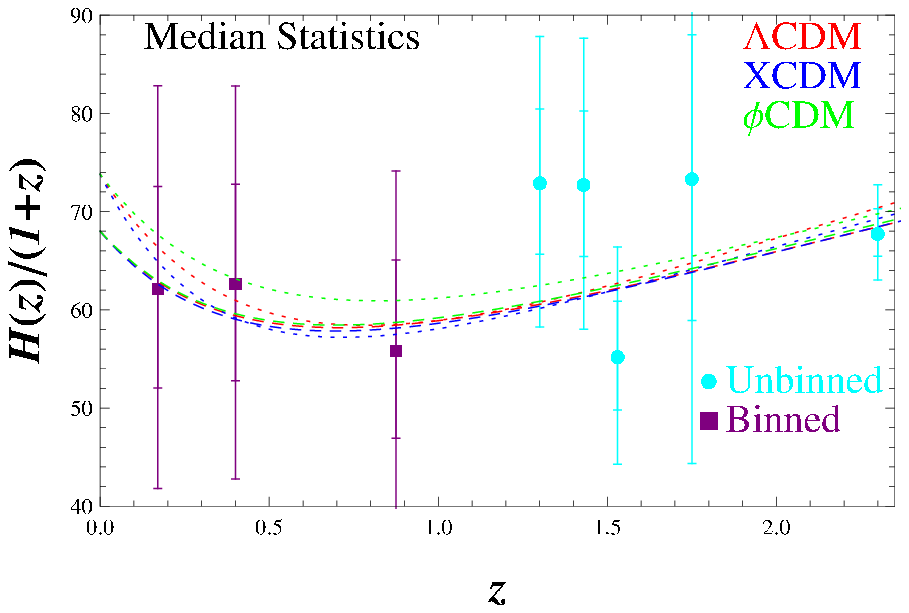}
  \includegraphics[width=39.5mm]{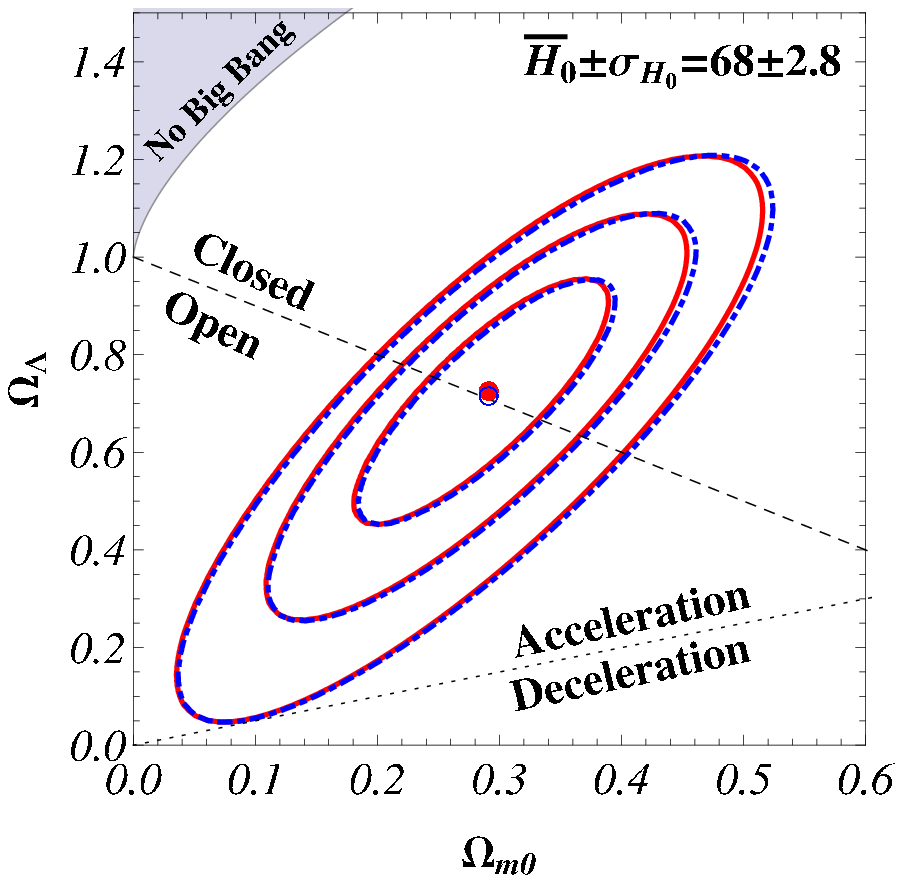}
  \includegraphics[width=39.5mm]{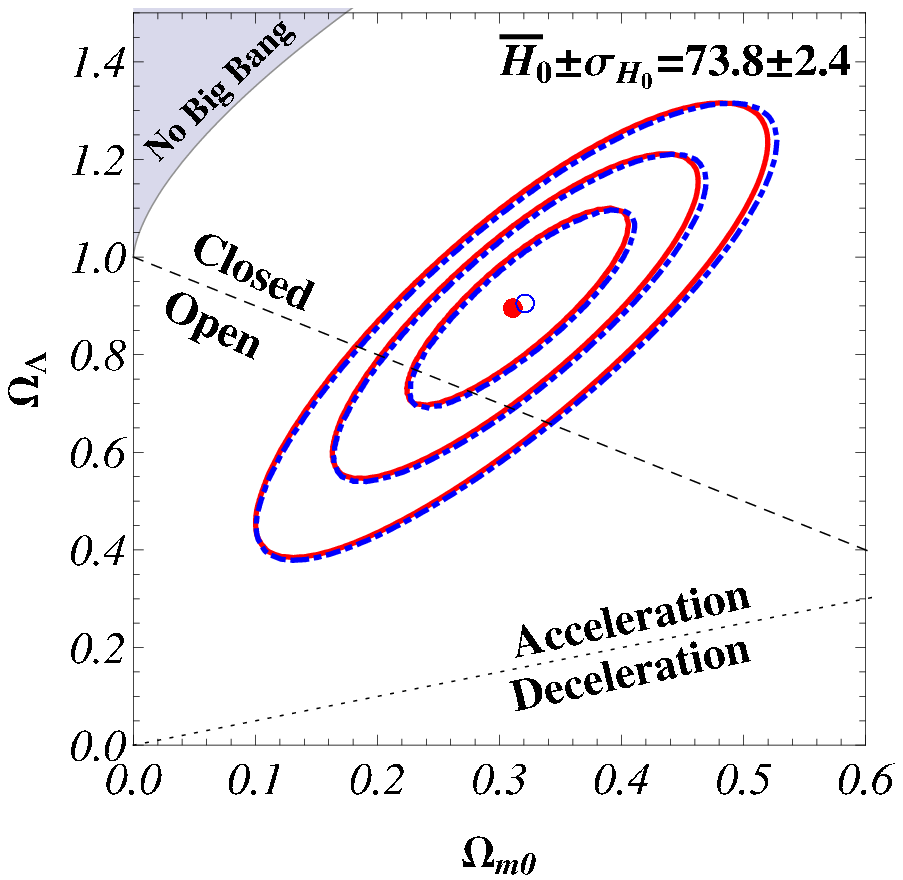}
  \includegraphics[width=39.5mm]{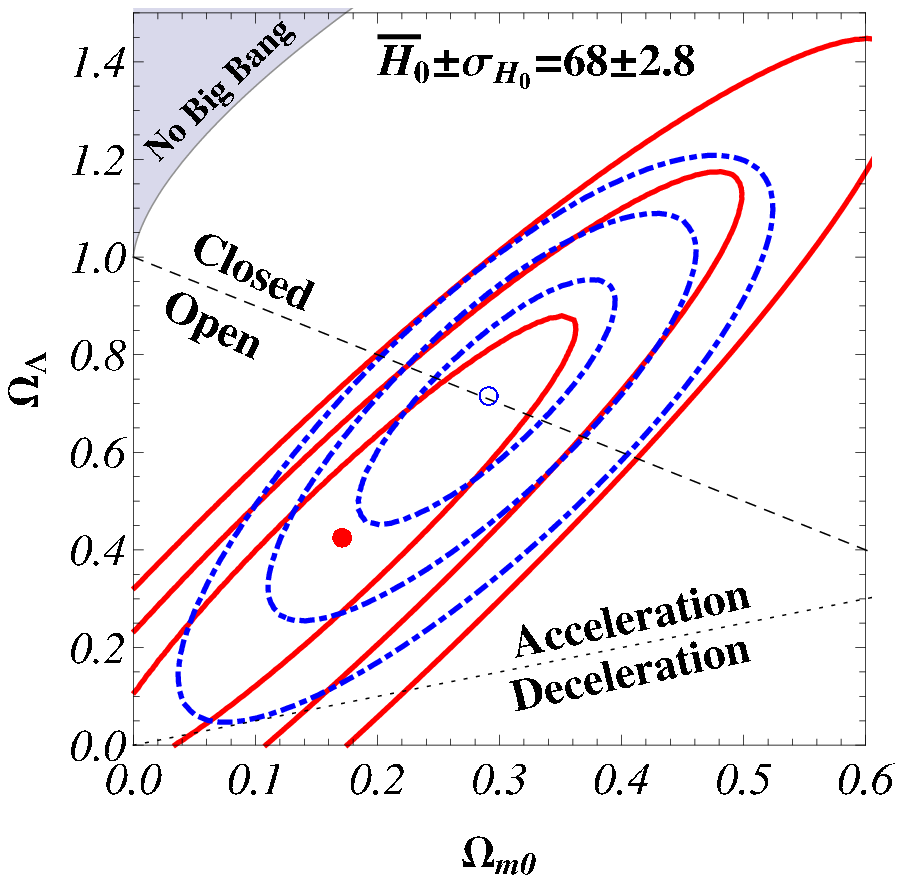}
  \includegraphics[width=39.5mm]{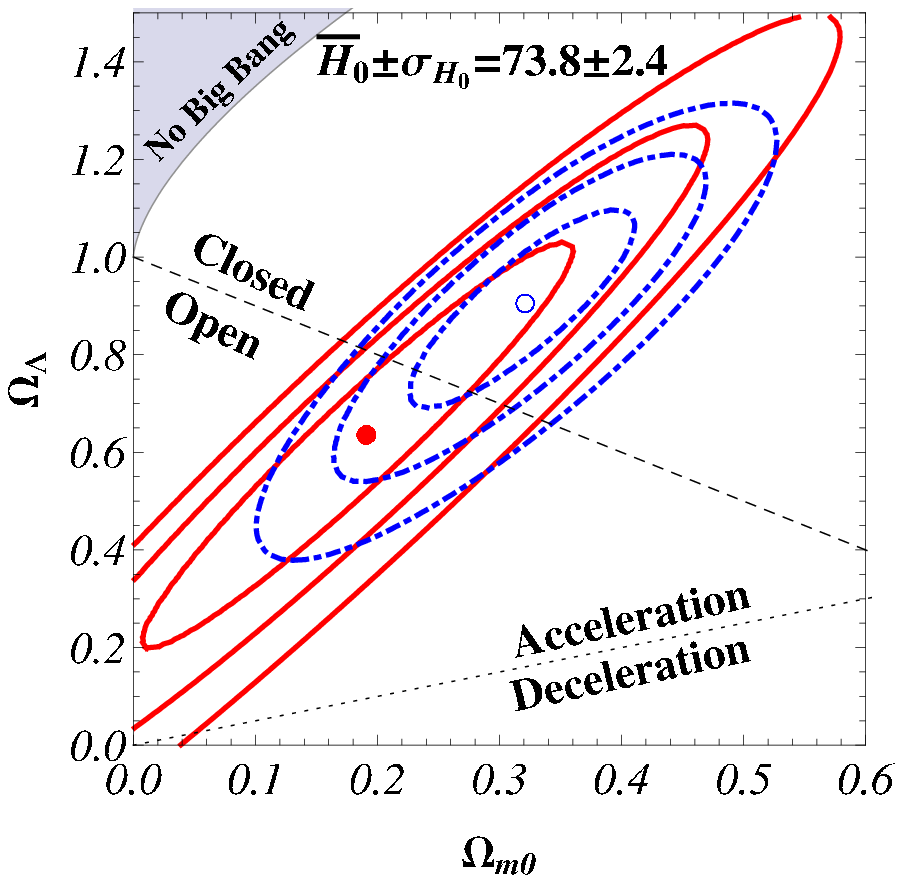}
  \includegraphics[width=40mm]{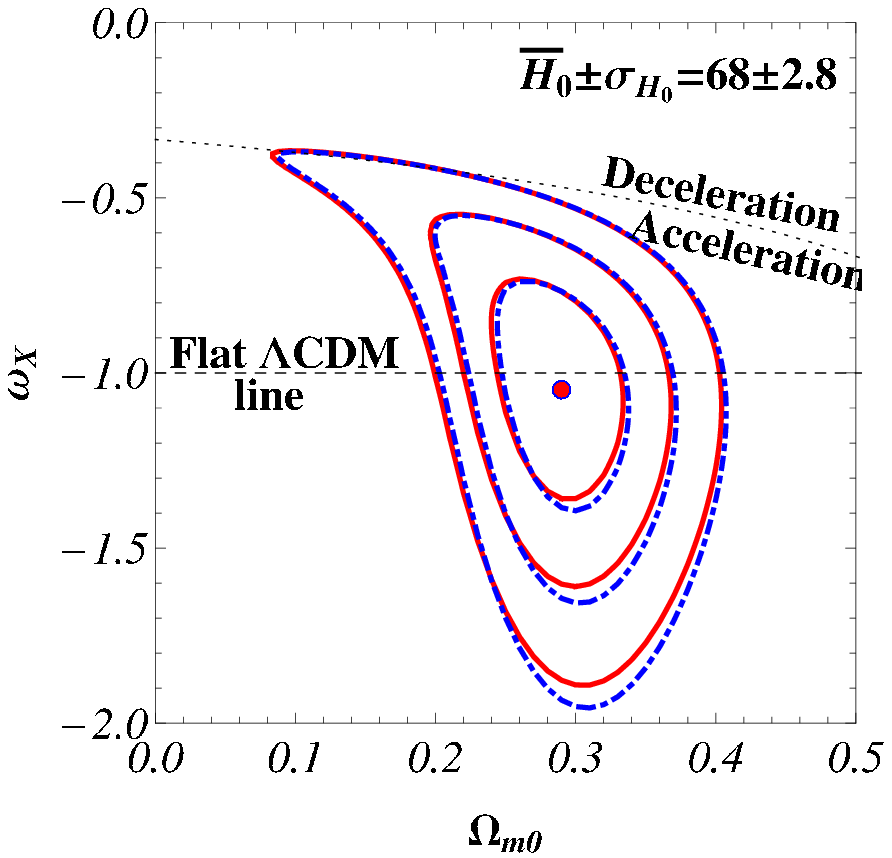}
  \includegraphics[width=40mm]{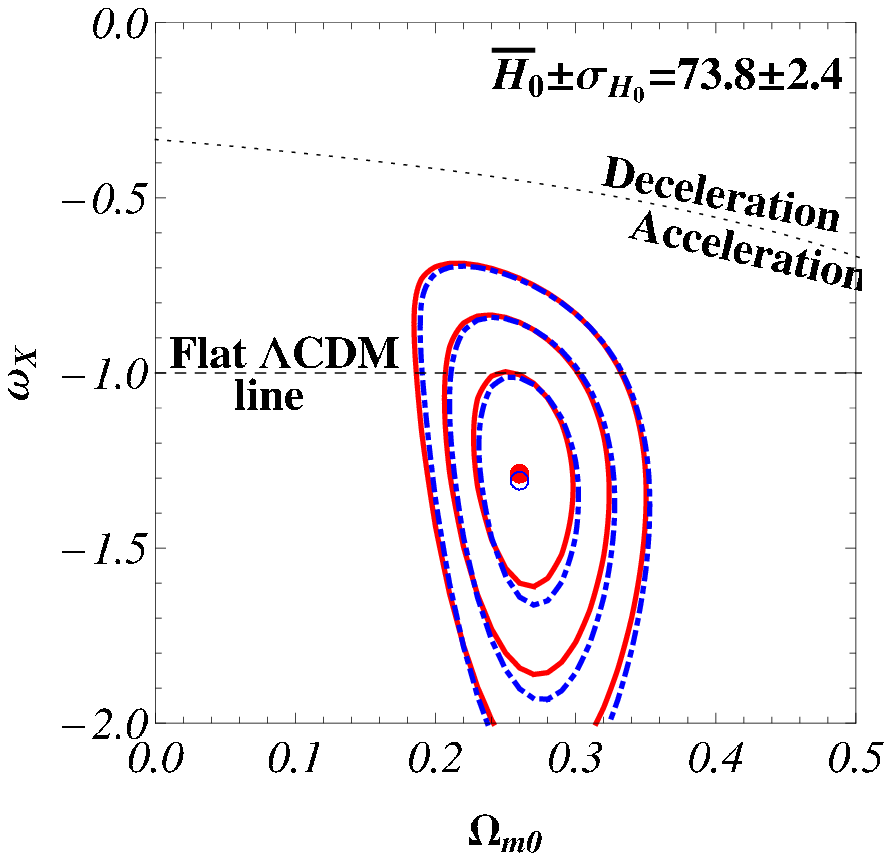}
  \includegraphics[width=40mm]{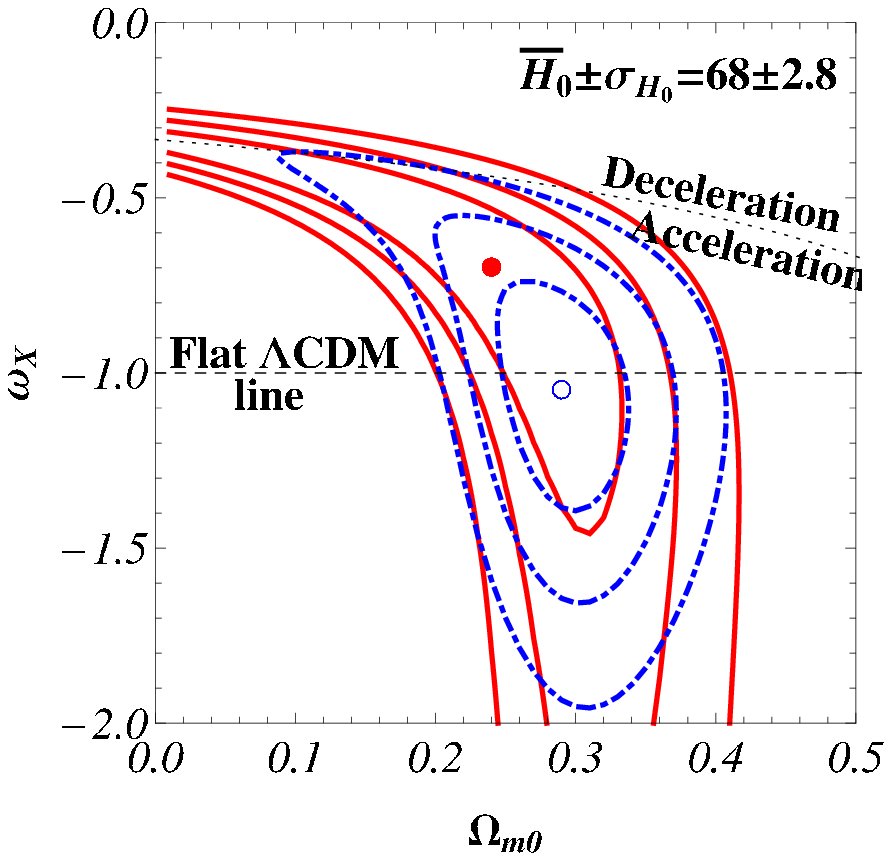}
  \includegraphics[width=40mm]{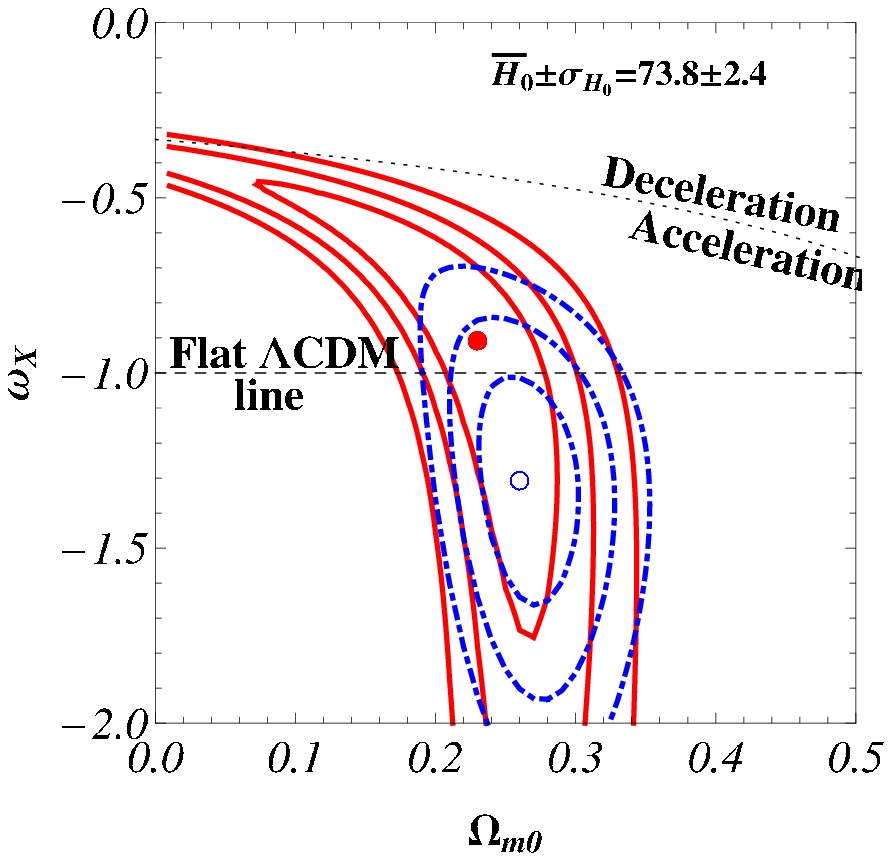}
  \includegraphics[width=39.5mm]{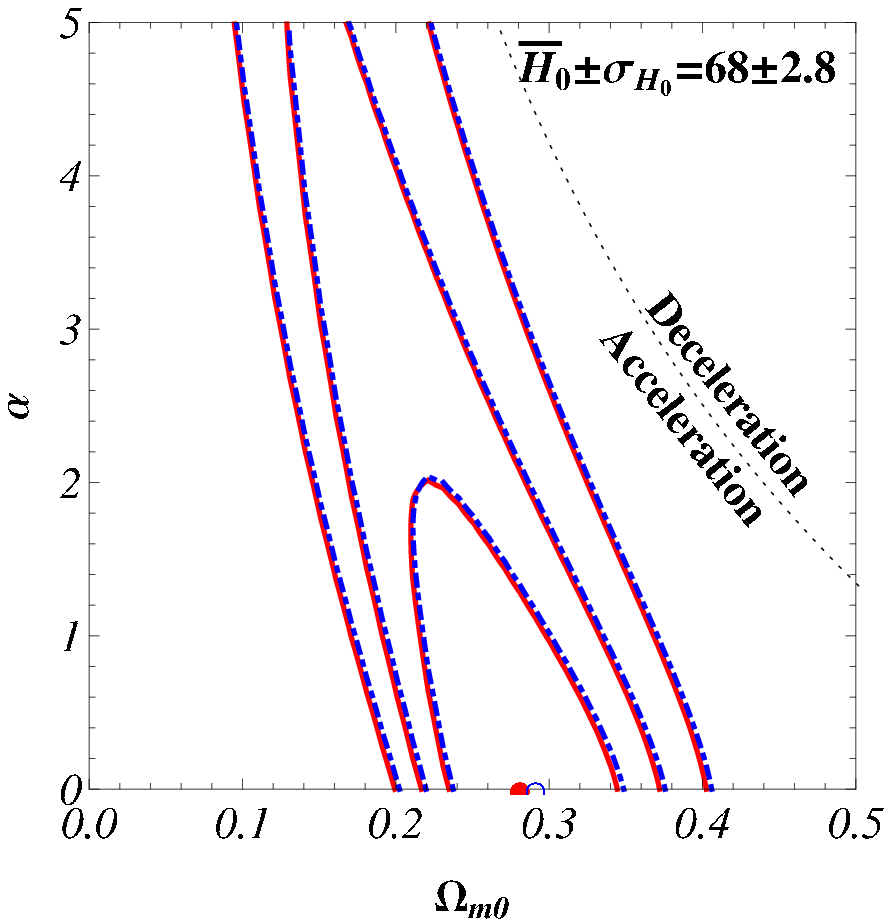}
  \includegraphics[width=39.5mm]{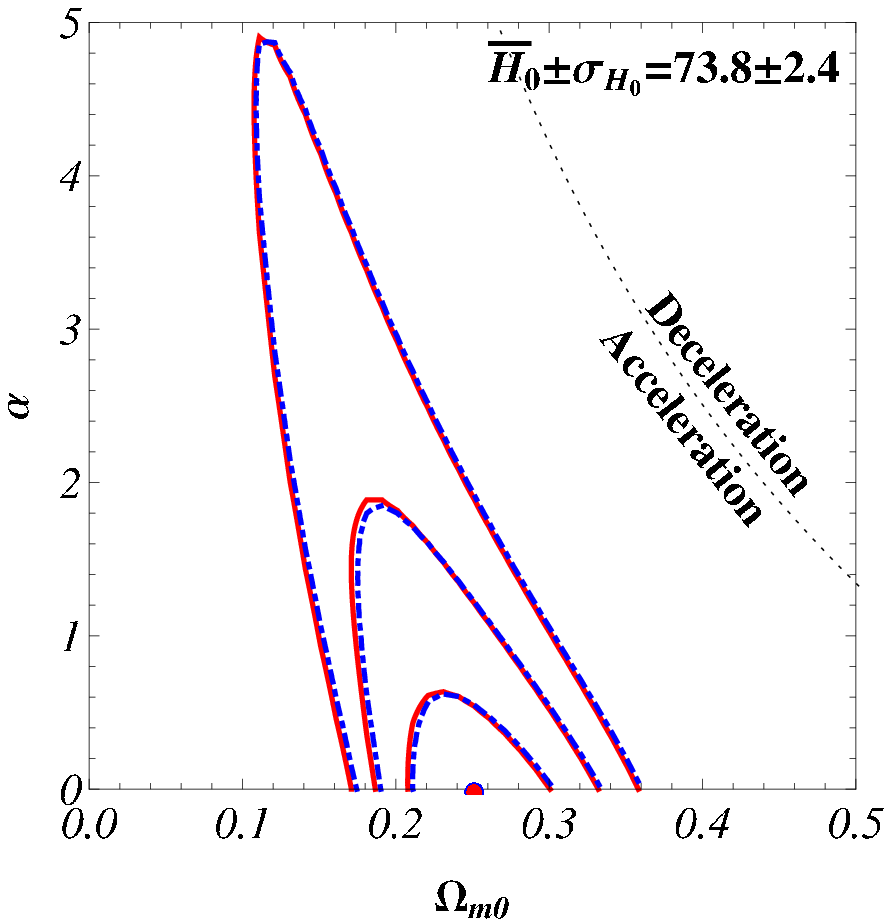}
  \includegraphics[width=39.5mm]{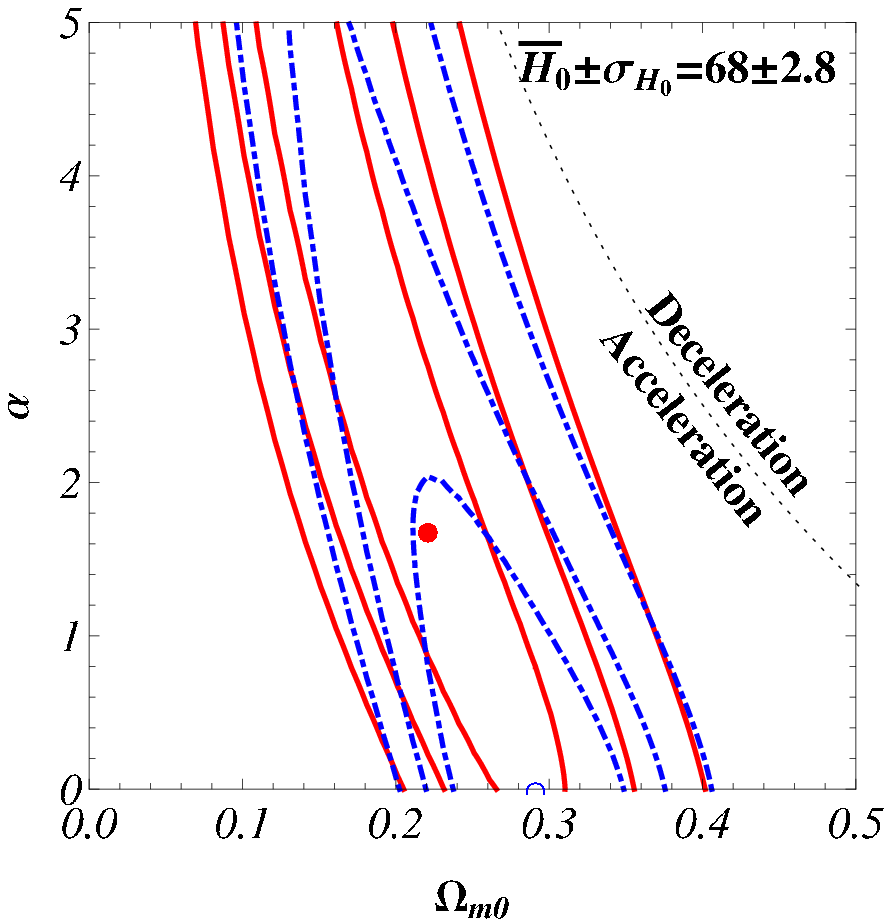}
  \includegraphics[width=39.5mm]{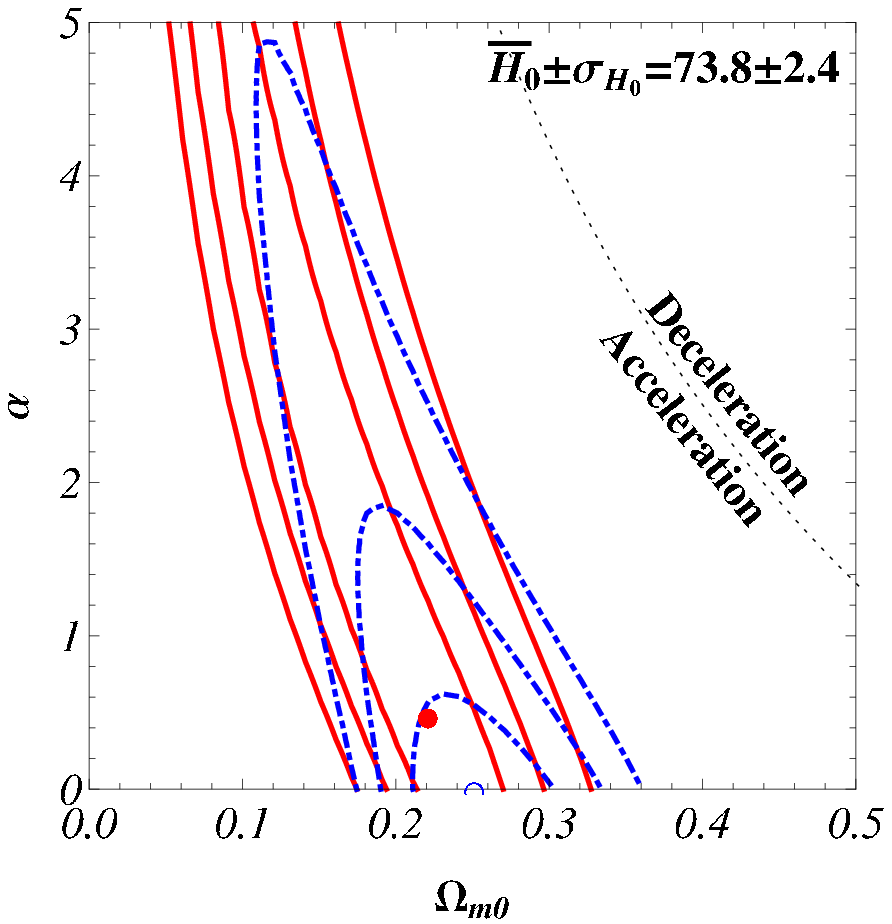}
\caption{
Top left (right) panel shows the $H(z)/(1+z)$ data, binned with 7 or 9 
measurements per bin,  as well as 5 higher $z$ measurements, and the FR 
best-fit model predictions, dashed (dotted) for lower (higher) $H_0$ prior. 
The 2nd through 4th rows show the $H(z)$ constraints for $\Lambda$CDM, 
XCDM, and $\phi$CDM.
Red (blue dot-dashed) contours are 1, 2, and 3 $\sigma$ confidence interval results from 7 or 9
measurements per bin (unbinned FR Table\ 1) data. 
In these three rows, the first two plots 
include red weighted-mean constraints while the second two include red median statistics ones. The 
filled red (empty blue) circle is the corresponding best-fit point.
Dashed diagonal lines show spatially-flat models, and dotted
lines indicate zero-acceleration models. For quantitative details see Table \ref{table:fig4 details}.
} \label{fig:For table 8,9}
\end{figure}

\newpage 

%%%%%%%%%%%%
%%% Table 1.
%%%%%%%%%%%%

\begin{deluxetable}{ccccc}
\tablecaption{Deceleration-Acceleration Transition Redshifts\tablenotemark{a}}
\tablewidth{0pt}
%\tabletypesize{\small}
\tablehead{ 
\colhead{$h$ Prior\tablenotemark{b}}& 
\colhead{Best-Fit Values}& 
\colhead{$\chi_{\rm min}^2$}&
\colhead{$z_{da} \pm \sigma_{z_{da}}$\tablenotemark{c}} &
\colhead{$z_{da}\tablenotemark{d}$}\\
\noalign{\vskip -4mm}
}
\startdata
\noalign{\vskip -0mm}
\multicolumn{5}{c}{$\Lambda$CDM}\\
\noalign{\vskip 1mm}
\hline
%\cutinhead{$\Lambda$CDM}
\noalign{\vskip 1mm} 
0.68 $\pm$ 0.028&	$(\Omega_{m0},\Omega_{\Lambda})=(0.29,0.72)$&	18.2&	0.690 $\pm$ 0.096&	   0.706\\
\noalign{\vskip 2mm}
0.738 $\pm$ 0.024& $(\Omega_{m0},\Omega_{\Lambda})=(0.32,0.91$)&	19.3&	0.781 $\pm$ 0.067&	   0.785\\

\cutinhead{XCDM}
0.68 $\pm$ 0.028&	$(\Omega_{m0},\omega_{X})=(0.29,-1.04)$&	18.2&	0.677 $\pm$ 0.097&	   0.695\\
\noalign{\vskip 2mm}
0.738 $\pm$ 0.024&	 $(\Omega_{m0},\omega_{X})=(0.26,-1.30)$&	18.2&	0.696 $\pm$ 0.082&	   0.718\\

\cutinhead{$\phi$CDM}
0.68 $\pm$ 0.028&	$(\Omega_{m0},\alpha)=(0.29,0.00)$&	18.2&	0.724 $\pm$ 0.148&	   0.698\\
\noalign{\vskip 2mm}
0.738 $\pm$ 0.024&	 $(\Omega_{m0},\alpha)=(0.25,0.00)$&	20.7&	0.850 $\pm$ 0.116&	   0.817\\

\enddata
\tablenotetext{a}{Estimated using the unbinned data in Table 1 of FR.}
\tablenotetext{b}{Hubble constant in units of 100 km s$^{-1}$ Mpc$^{-1}$.}
\tablenotetext{c}{Computed using Eqs. (1-6).} 
\tablenotetext{d}{The deceleration-acceleration transition redshift in the model
with the best-fit values of the cosmological parameters, as computed in FR.} 
\label{table:Un-binned data details}
\end{deluxetable}

%%%%%%%%%%%%%%%%
%%% Table 1.End
%%%%%%%%%%%%%%%%

\newpage

%%%%%%%%%%%%
%%% Table 2.
%%%%%%%%%%%%

\renewcommand{\arraystretch}{1.1}
\begin{deluxetable}{lcccccc}
\tablecaption{Weighted Mean Results For 23 Lower Redshift Measurements}
\tablewidth{0pt}
\tabletypesize{\small}
\tablehead{
\colhead{\multirow{2}{*}{Bin}}& 
\colhead{\multirow{2}{*}{$N$}}& 
\colhead{\multirow{2}{*}{$z^a$}}& 
\colhead{$H(z)$}&
\colhead{$H(z)$ (1 $\sigma$ range)} &
\colhead{$H(z)$ (2 $\sigma$ range)}&
\colhead{\multirow{2}{*}{$N_\sigma$}}\\
&
&
&
(km s$^{-1}$ Mpc $^{-1}$)&
(km s$^{-1}$ Mpc $^{-1}$)&
(km s$^{-1}$ Mpc $^{-1}$)&
}
\startdata
\noalign{\vskip -1mm}
\multicolumn{7}{c}{3 or 4 measurements per bin}\\
\noalign{\vskip 1mm}
\hline
%\cutinhead{3 or 4 measurements per bin}
\noalign{\vskip 2mm} 
1&	3&	0.096&	69.0&	59.4$-$78.5&   49.9$-$88.0&	2.00\\

2&	4&	0.185&	76.0&	73.1$-$78.9&   70.2$-$81.8&	1.73\\

3&	3&  	0.338&	76.6&	71.5$-$81.8&   66.4$-$86.9&   1.89\\

4&	3&	0.417&	84.4&	78.1$-$90.7&	71.8$-$97.0&	1.55\\

5&	3&	0.598&	90.9&	85.4$-$96.4&	  79.9$-$102&	0.73\\

6&	3&	0.720&	96.6&         91.8$-$101&  87.0$-$106&	1.17\\

7&	4&	0.929&	129&	118$-$140&   107$-$151&	0.13\\

\cutinhead{4 or 5 measurements per bin}
1&	4&	0.139&	77.2&	71.1$-$83.3&   64.9$-$89.5&	1.41\\

2&	5&	0.191&	75.2&	72.1$-$78.2&   69.1$-$81.2&	2.71\\

3&	5&      0.380&	79.9&	75.8$-$84.1&   71.6$-$88.3&  1.81\\

4&	5&	0.668&	94.1&	90.5$-$97.7&   86.8$-$101&	0.91\\

5&	4&	0.929&	129&	118$-$140&   107$-$151&  0.13\\

\cutinhead{5 or 6 measurements per bin}
1&	5&	0.167&	75.7&	72.3$-$79.0&   69.0$-$82.3&	1.86\\

2&	6&	0.271&	76.2&	72.7$-$79.7&   69.3$-$83.1&	2.89\\

3&	6&      0.569&	89.4&	85.5$-$93.2&   81.7$-$97.0&   1.70\\

4&	6&	0.787&	106&	101$-$112&   95.8$-$117&	2.50\\
\cutinhead{7 or 9 measurements per bin}
1&	7&	0.177&	75.4&	72.7$-$78.2&   69.9$-$81.0&	2.66\\

2&	9&	0.448&	83.6&	80.3$-$86.8&   77.1$-$90.0&	1.29\\

3&	7&	0.754&	102&	97.6$-$106&   93.2$-$111&   2.99
\enddata
\tablenotetext{a}{Weighted mean of $z$ values of measurements in the bin.} 
\label{table:WA}
\end{deluxetable}
%%%%%%%%%%%%%%%%
%%% Table 2.End
%%%%%%%%%%%%%%%%

\newpage

%%%%%%%%%%%%
%%% Table 3.
%%%%%%%%%%%%
\renewcommand{\arraystretch}{1.1}
\begin{deluxetable}{lccccc}
\tablecaption{Median Statistics Results For 23 Lower Redshift Measurements}
\tablewidth{0pt}
\tabletypesize{\small}
\tablehead{
\colhead{\multirow{2}{*}{Bin}}& 
\colhead{\multirow{2}{*}{$N$}}& 
\colhead{\multirow{2}{*}{$z^b$}}& 
\colhead{$H(z)$}&
\colhead{$H(z)$ (1 $\sigma$ range)} &
\colhead{$H(z)$ (2 $\sigma$ range)}\\
&
&
&
(km s$^{-1}$ Mpc $^{-1}$)&
(km s$^{-1}$ Mpc $^{-1}$)&
(km s$^{-1}$ Mpc $^{-1}$)
}
\startdata
\noalign{\vskip -1mm}
\multicolumn{6}{c}{3 or 4 measurements per bin}\\
\noalign{\vskip 1mm}
\hline
%\cutinhead{3 or 4 measurements per bin}
\noalign{\vskip 2mm} 

1&	3&	0.100&	69.0&	49.4$-$88.6&   29.8$-$108\\

2&	4&	0.189&	75.0&	68.5$-$81.5&    62.0$-$88.0\\

3&	3&      0.280&	77.0&	63.0$-$91.0&   49.0$-$105\\

4&	3&	0.400&	83.0&	69.0$-$97.0&   55.0$-$111\\

5&	3&	0.593&	97.0&	84.0$-$110&   71.0$-$123\\

6&	3&	0.730&	97.3&  89.3$-$105&  81.3$-$113\\

7&	4&	0.890&	121&	99.5$-$143&  78.0$-$164\\

\cutinhead{4 or 5 measurements per bin}

1&	4&	0.110&	69.0&	53.2$-$84.8&    37.4$-$101\\

2&	5&	0.200&	75.0&	61.0$-$89.0&    47.0$-$103\\

3&	5&      0.400&	83.0&	69.0$-$97.0&    55.0$-$111\\

4&	5&	0.680&	97.3&	89.3$-$105&   81.3$-$113\\

5&	4&	0.890&	121&	99.5$-$143&    78.0$-$164\\

\cutinhead{5 or 6 measurements per bin}

1&	5&	0.120&	69.0&	57.0$-$81.0&   45.0$-$93.0\\

2&	6&	0.275&	76.7&	62.7$-$90.7&   48.7$-$105\\

3&	6&      0.537&	93.5&	83.0$-$104&   72.5$-$115\\

4&	6&	0.878&	111&	92.5$-$130&   74.0$-$148\\

\cutinhead{7 or 9 measurements per bin}

1&	7&	0.170&	72.9&	60.9$-$84.9&   48.9$-$96.9\\

2&	9&	0.400&	87.9&	73.9$-$102&   59.9$-$116\\

3&	7&	0.875&	105&	88.0$-$122&   71.0$-$139

\enddata
\tablenotetext{b}{Median of $z$ values of measurements in the bin.}
\label{table:Med}
\end{deluxetable}

%%%%%%%%%%%%%%%%
%%% Table 3.End
%%%%%%%%%%%%%%%%

%\newpage

%%%%%%%%%%%%%%%%%%%%%%%%%%%%
%Table 4
%%%%%%%%%%%%%%%%%%%%%%%%%%%%%%
\begin{table}[t]
\centering
\caption{Best-Fit Points And Minimum $\chi^2$s For 3 Or 4 Measurements Per Bin}
\vspace{4 mm}
\begin{tabular}{c c c c c c} 
\hline\hline 
{} & {} & \multicolumn{2}{c}{Weighted Mean} & \multicolumn{2}{c}{Median}\\

Model & $h$ Prior &  BFP & $\chi^2_{\rm min}$ &  BFP & $\chi^2_{\rm min}$\\

\hline

\multirow{4}{*}{$\Lambda$CDM} & \multirow{2}{*}{$0.68 \pm 0.028$} & $\Omega_{m0}=0.29$ &\multirow{2}{*}{13.0} & $\Omega_{m0}=0.24$ & \multirow{2}{*}{8.75}\\

& {} & $\Omega_{\Lambda}=0.72$ &{} & $\Omega_{\Lambda}=0.60$ \\

\cline{2-6}

& \multirow{2}{*}{$0.738 \pm 0.024$} & $\Omega_{m0}=0.32$ & \multirow{2}{*}{ 14.1} & $\Omega_{m0}=0.26$ & \multirow{2}{*}{9.37}\\

&{}& $\Omega_{\Lambda}=0.91$ &{}& $\Omega_{\Lambda}=0.80$\\ 

\hline

\multirow{4}{*}{XCDM} & \multirow{2}{*}{$0.68 \pm 0.028$} & $\Omega_{m0}=0.29$ &\multirow{2}{*}{13.0} & $\Omega_{m0}=0.28$ & \multirow{2}{*}{8.85}\\

& {} & $\omega_{X}=-1.04$ &{} & $\omega_{X}=-0.90$ \\

\cline{2-6}

& \multirow{2}{*}{$0.738 \pm 0.024$} & $\Omega_{m0}=0.26$ & \multirow{2}{*}{ 13.0} & $\Omega_{m0}=0.25$ & \multirow{2}{*}{9.18}\\

&{}& $\omega_{X}=-1.29$ &{}& $\omega_{X}=-1.13$\\ 

\hline

\multirow{4}{*}{$\phi$CDM} & \multirow{2}{*}{$0.68 \pm 0.028$} & $\Omega_{m0}=0.29$ &\multirow{2}{*}{13.0} & $\Omega_{m0}=0.27$ & \multirow{2}{*}{8.82}\\

& {} & $\alpha=0.00$ &{} & $\alpha=0.46$ \\

\cline{2-6}

& \multirow{2}{*}{$0.738 \pm 0.024$} & $\Omega_{m0}=0.25$ & \multirow{2}{*}{ 15.4} & $\Omega_{m0}=0.24$ & \multirow{2}{*}{9.43}\\

&{}& $\alpha=0.00$ &{}& $\alpha=0.00$\\ 

\hline  
\hline
\end{tabular}
\label{table:fig1 details}
\end{table}
\vspace{-8 mm}
%%%%%%%%%%%%%%%%%%%%%%%%%%%%
%Table 5
%%%%%%%%%%%%%%%%%%%%%%%%%%%%%%
\begin{table}[t]
\centering
\caption{Best-Fit Points And Minimum $\chi^2$s For 4 Or 5 Measurements Per Bin}
\vspace{4 mm}
\begin{tabular}{c c c c c c} 
\hline\hline

{} & {} & \multicolumn{2}{c}{Weighted Mean} & \multicolumn{2}{c}{Median}\\

Model & $h$ Prior &  BFP & $\chi^2_{\rm min}$ &  BFP & $\chi^2_{\rm min}$\\

\hline

\multirow{4}{*}{$\Lambda$CDM} & \multirow{2}{*}{$0.68 \pm 0.028$} & $\Omega_{m0}=0.29$ &\multirow{2}{*}{12.9} & $\Omega_{m0}=0.17$ & \multirow{2}{*}{7.62}\\

& {} & $\Omega_{\Lambda}=0.73$ &{} & $\Omega_{\Lambda}=0.43$ \\

\cline{2-6}

& \multirow{2}{*}{$0.738 \pm 0.024$} & $\Omega_{m0}=0.32$ & \multirow{2}{*}{ 13.7} & $\Omega_{m0}=0.20$ & \multirow{2}{*}{8.04}\\

&{}& $\Omega_{\Lambda}=0.91$ &{}& $\Omega_{\Lambda}=0.66$\\ 

\hline

\multirow{4}{*}{XCDM} & \multirow{2}{*}{$0.68 \pm 0.028$} & $\Omega_{m0}=0.29$ &\multirow{2}{*}{12.9} & $\Omega_{m0}=0.24$ & \multirow{2}{*}{7.75}\\

& {} & $\omega_{X}=-1.06$ &{} & $\omega_{X}=-0.68$ \\

\cline{2-6}

& \multirow{2}{*}{$0.738 \pm 0.024$} & $\Omega_{m0}=0.26$ & \multirow{2}{*}{ 12.5} & $\Omega_{m0}=0.23$ & \multirow{2}{*}{8.17}\\

&{}& $\omega_{X}=-1.31$ &{}& $\omega_{X}=-0.89$\\ 

\hline

\multirow{4}{*}{$\phi$CDM} & \multirow{2}{*}{$0.68 \pm 0.028$} & $\Omega_{m0}=0.29$ &\multirow{2}{*}{13.0} & $\Omega_{m0}=0.22$ & \multirow{2}{*}{7.70}\\

& {} & $\alpha=0.00$ &{} & $\alpha=1.77$ \\

\cline{2-6}

& \multirow{2}{*}{$0.738 \pm 0.024$} & $\Omega_{m0}=0.25$ & \multirow{2}{*}{ 15.2} & $\Omega_{m0}=0.23$ & \multirow{2}{*}{8.13}\\

&{}& $\alpha=0.00$ &{}& $\alpha=0.30$\\ 

\hline  
\hline
\end{tabular}
\label{table:fig2 details}
\end{table}

\vspace{-8 mm}
%%%%%%%%%%%%%%%%%%%%%%%%%%%%
%Table 6
%%%%%%%%%%%%%%%%%%%%%%%%%%%%%%
\begin{table}[t]
\centering
\caption{Best-Fit Points And Minimum $\chi^2$s For 5 Or 6 Measurements Per Bin}
\vspace{4 mm}
\begin{tabular}{c c c c c c} 
\hline\hline 

{} & {} & \multicolumn{2}{c}{Weighted Mean} & \multicolumn{2}{c}{Median}\\

Model & $h$ Prior &  BFP & $\chi^2_{\rm min}$ &  BFP & $\chi^2_{\rm min}$\\

\hline

\multirow{4}{*}{$\Lambda$CDM} & \multirow{2}{*}{$0.68 \pm 0.028$} & $\Omega_{m0}=0.28$ &\multirow{2}{*}{10.2} & $\Omega_{m0}=0.18$ & \multirow{2}{*}{7.65}\\

& {} & $\Omega_{\Lambda}=0.70$ &{} & $\Omega_{\Lambda}=0.45$ \\

\cline{2-6}

& \multirow{2}{*}{$0.738 \pm 0.024$} & $\Omega_{m0}=0.31$ & \multirow{2}{*}{ 11.2} & $\Omega_{m0}=0.20$ & \multirow{2}{*}{8.13}\\

&{}& $\Omega_{\Lambda}=0.89$ &{}& $\Omega_{\Lambda}=0.66$\\ 

\hline

\multirow{4}{*}{XCDM} & \multirow{2}{*}{$0.68 \pm 0.028$} & $\Omega_{m0}=0.29$ &\multirow{2}{*}{10.2} & $\Omega_{m0}=0.24$ & \multirow{2}{*}{7.77}\\

& {} & $\omega_{X}=-1.01$ &{} & $\omega_{X}=-0.68$ \\

\cline{2-6}

& \multirow{2}{*}{$0.738 \pm 0.024$} & $\Omega_{m0}=0.26$ & \multirow{2}{*}{ 10.2} & $\Omega_{m0}=0.23$ & \multirow{2}{*}{8.25}\\

&{}& $\omega_{X}=-1.28$ &{}& $\omega_{X}=-0.89$\\ 

\hline

\multirow{4}{*}{$\phi$CDM} & \multirow{2}{*}{$0.68 \pm 0.028$} & $\Omega_{m0}=0.29$ &\multirow{2}{*}{10.2} & $\Omega_{m0}=0.22$ & \multirow{2}{*}{7.72}\\

& {} & $\alpha=0.00$ &{} & $\alpha=1.73$ \\

\cline{2-6}

& \multirow{2}{*}{$0.738 \pm 0.024$} & $\Omega_{m0}=0.25$ & \multirow{2}{*}{ 12.4} & $\Omega_{m0}=0.23$ & \multirow{2}{*}{8.21}\\

&{}& $\alpha=0.00$ &{}& $\alpha=0.30$\\ 

\hline  
\hline
\end{tabular}
\label{table:fig3 details}
\end{table}
\vspace{-8 mm}
%%%%%%%%%%%%%%%%%%%%%%%%%%%%
%Table 7
%%%%%%%%%%%%%%%%%%%%%%%%%%%%%%
\begin{table}[t]
\centering
\caption{Best-Fit Points And Minimum $\chi^2$s For 7 Or 9 Measurements Per Bin}
\vspace{4 mm}
\begin{tabular}{c c c c c c} 
\hline\hline 

{} & {} & \multicolumn{2}{c}{Weighted Mean} & \multicolumn{2}{c}{Median}\\

Model & $h$ Prior &  BFP & $\chi^2_{\rm min}$ &  BFP & $\chi^2_{\rm min}$\\

\hline

\multirow{4}{*}{$\Lambda$CDM} & \multirow{2}{*}{$0.68 \pm 0.028$} & $\Omega_{m0}=0.29$ &\multirow{2}{*}{9.7} & $\Omega_{m0}=0.17$ & \multirow{2}{*}{7.76}\\

& {} & $\Omega_{\Lambda}=0.73$ &{} & $\Omega_{\Lambda}=0.43$ \\

\cline{2-6}

& \multirow{2}{*}{$0.738 \pm 0.024$} & $\Omega_{m0}=0.31$ & \multirow{2}{*}{ 10.4} & $\Omega_{m0}=0.19$ & \multirow{2}{*}{7.88}\\

&{}& $\Omega_{\Lambda}=0.90$ &{}& $\Omega_{\Lambda}=0.64$\\ 

\hline

\multirow{4}{*}{XCDM} & \multirow{2}{*}{$0.68 \pm 0.028$} & $\Omega_{m0}=0.29$ &\multirow{2}{*}{9.7} & $\Omega_{m0}=0.24$ & \multirow{2}{*}{7.82}\\

& {} & $\omega_{X}=-1.04$ &{} & $\omega_{X}=-0.69$ \\

\cline{2-6}

& \multirow{2}{*}{$0.738 \pm 0.024$} & $\Omega_{m0}=0.26$ & \multirow{2}{*}{ 9.5} & $\Omega_{m0}=0.23$ & \multirow{2}{*}{7.97}\\

&{}& $\omega_{X}=-1.28$ &{}& $\omega_{X}=-0.90$\\ 

\hline

\multirow{4}{*}{$\phi$CDM} & \multirow{2}{*}{$0.68 \pm 0.028$} & $\Omega_{m0}=0.28$ &\multirow{2}{*}{9.7} & $\Omega_{m0}=0.22$ & \multirow{2}{*}{7.80}\\

& {} & $\alpha=0.00$ &{} & $\alpha=1.69$ \\

\cline{2-6}

& \multirow{2}{*}{$0.738 \pm 0.024$} & $\Omega_{m0}=0.25$ & \multirow{2}{*}{ 11.8} & $\Omega_{m0}=0.22$ & \multirow{2}{*}{7.93}\\

&{}& $\alpha=0.00$ &{}& $\alpha=0.48$\\ 

\hline  
\hline
\end{tabular}
\label{table:fig4 details}
\end{table}

\end{document}